\PassOptionsToPackage{unicode}{hyperref}
\PassOptionsToPackage{hyphens}{url}
\documentclass[
]{article}
\usepackage{lmodern}
\usepackage{amssymb,amsmath}
\usepackage{ifxetex,ifluatex}
\ifnum 0\ifxetex 1\fi\ifluatex 1\fi=0 % if pdftex
  \usepackage[T1]{fontenc}
  \usepackage[utf8]{inputenc}
  \usepackage{textcomp} % provide euro and other symbols
\else % if luatex or xetex
  \usepackage{unicode-math}
  \defaultfontfeatures{Scale=MatchLowercase}
  \defaultfontfeatures[\rmfamily]{Ligatures=TeX,Scale=1}
\fi
\IfFileExists{upquote.sty}{\usepackage{upquote}}{}
\IfFileExists{microtype.sty}{% use microtype if available
  \usepackage[]{microtype}
  \UseMicrotypeSet[protrusion]{basicmath} % disable protrusion for tt fonts
}{}
\makeatletter
\@ifundefined{KOMAClassName}{% if non-KOMA class
  \IfFileExists{parskip.sty}{%
    \usepackage{parskip}
  }{% else
    \setlength{\parindent}{0pt}
    \setlength{\parskip}{6pt plus 2pt minus 1pt}}
}{% if KOMA class
  \KOMAoptions{parskip=half}}
\makeatother
\usepackage{xcolor}
\IfFileExists{xurl.sty}{\usepackage{xurl}}{} % add URL line breaks if available
\IfFileExists{bookmark.sty}{\usepackage{bookmark}}{\usepackage{hyperref}}
\hypersetup{
  pdftitle={Abstract Orientable Incidence Structure and Algorithms for Finite Bounded Acyclic Categories. II. Data Structure and Fundamental Operations},
  pdfauthor={Yu-Wei Huang},
  hidelinks,
  pdfcreator={LaTeX via pandoc}}
\usepackage{color}
\usepackage{fancyvrb}

\newcommand{\VERB}{\Verb[commandchars=\\\{\}]}
\DefineVerbatimEnvironment{Highlighting}{Verbatim}{commandchars=\\\{\}}
\usepackage{framed}
\definecolor{shadecolor}{RGB}{248,248,248}
\newenvironment{Shaded}{\begin{snugshade}}{\end{snugshade}}

\newcommand{\CommentTok}[1]{\textcolor[rgb]{0.56,0.35,0.01}{\textit{#1}}}

\newcommand{\DataTypeTok}[1]{\textcolor[rgb]{0.13,0.29,0.53}{#1}}

\newcommand{\FunctionTok}[1]{\textcolor[rgb]{0.00,0.00,0.00}{#1}}

\newcommand{\KeywordTok}[1]{\textcolor[rgb]{0.13,0.29,0.53}{\textbf{#1}}}
\newcommand{\NormalTok}[1]{#1}
\newcommand{\OperatorTok}[1]{\textcolor[rgb]{0.81,0.36,0.00}{\textbf{#1}}}
\newcommand{\OtherTok}[1]{\textcolor[rgb]{0.56,0.35,0.01}{#1}}

\usepackage{graphicx}
\makeatletter
\def\maxwidth{\ifdim\Gin@nat@width>\linewidth\linewidth\else\Gin@nat@width\fi}
\def\maxheight{\ifdim\Gin@nat@height>\textheight\textheight\else\Gin@nat@height\fi}
\makeatother
\setkeys{Gin}{width=\maxwidth,height=\maxheight,keepaspectratio}
\makeatletter
\def\fps@figure{htbp}
\makeatother
\providecommand{\tightlist}{%
  \setlength{\itemsep}{0pt}\setlength{\parskip}{0pt}}
\usepackage{fullpage}
\usepackage[]{biblatex}
\title{Abstract Orientable Incidence Structure and Algorithms for Finite
Bounded Acyclic Categories. II. Data Structure and Fundamental
Operations}
\author{Yu-Wei Huang\thanks{l28071504@gs.ncku.edu.tw}}
\date{\today}

\begin{document}
\maketitle
\begin{abstract}
A data structure for finite bounded acyclic categories has been built,
which is useful to encode and manipulate abstract orientable incidence
structure. It can be represented as a directed acyclic multigraph with
weighted edges, where the weighs encode the algebraic structure between
edges. The fundamental operations on this data structure are
investigated from geometrical, categorical and programming perspectives.
\end{abstract}

\hypertarget{introduction}{%
\section{Introduction}\label{introduction}}

In the previous article\autocite{huang2023abstract}, we introduced the
orientable incidence structure, which is a bounded acyclic category with
some additional properties. It provides another approach to investigate
the computer graphics in any dimension. In order to apply the theory to
practical use, the computer needs to understand the bounded acyclic
category. It's more similar to dealing with graphs than categories; a
finite category can be treated as a transitive graph with algebraic
structure between edges. Moreover, manipulating finite bounded acyclic
categories is simpler than non-acyclic ones, just like manipulating
directed acyclic graphs is easier than general graphs. In this article,
an efficient data structure for finite bounded acyclic categories is
introduced, and fundamental operations on this data structure is
developed.

\hypertarget{data-structure}{%
\section{Data Structure}\label{data-structure}}

\hypertarget{model}{%
\subsection{Model}\label{model}}

The most intuitive way to encode bounded acyclic categories is using a
transitive directed multigraph, where nodes indicate objects, and edges
indicate morphisms. But this is not enough to describe a category, the
additional information required by the category is the composition of
morphisms. The computer should also store the multiplication table of
all morphisms, but this is a waste of memory because not all morphisms
are composable and some cases can be omitted due to transitivity. In the
previous article, we shown that a bounded acyclic category is equivalent
to the category of upper categories and downward functors, so one can
model the category of upper categories instead of the bounded acyclic
category, where upper categories are encoded as nodes, and downward
functors between them as edges. Since such category is acyclic, it forms
a transitive directed acyclic multigraph. Notice that there is exactly
one root node for this directed acyclic multigraph (representing the
host bounded acyclic category), so all others nodes (representing upper
categories) are directly connected from this root node by exactly one
edge (representing downward functors). There may have multiple edges
between nodes, and adjacent edges can be composed to an edge that
preserves transitivity, whose rule is determined by the downward
functors. That means storing all downward functors as edges is not
necessary, one can just keep some of them so that other downward
functors can be obtained by following along paths.

The objects of an upper category are encoded as a set of symbols on the
corresponding node, and object's mappings of downward functors between
upper categories are encoded as mappings between those symbols, which
are stored as the weights of edges. The downward functor represented by
a given path can be constructed by composing mappings along the path.
Objects and their mappings are now fully encoded. Morphisms and their
mappings are also full encoded by these information, as explained below.
There is no need to encode morphisms of upper categories: in an upper
category, initial morphisms are also represented by its objects, and
non-initial morphisms can be obtained by applying downward functors to
initial morphisms of some upper categories. There is also no need to
encode composition of morphisms: consider that a downward functor maps
an initial object to a non-initial object, then this downward functor
should also represent the initial morphism of this non-initial object.
So the mapping representing this downward functor also represents a
process of making composite morphism by the morphism this functor
indicated. At the end of the day, the computer just needs to remember
objects and the mappings between them.

\begin{figure}
\hypertarget{fig:utility-pole}{%
\centering
\includegraphics[width=1\textwidth,height=\textheight]{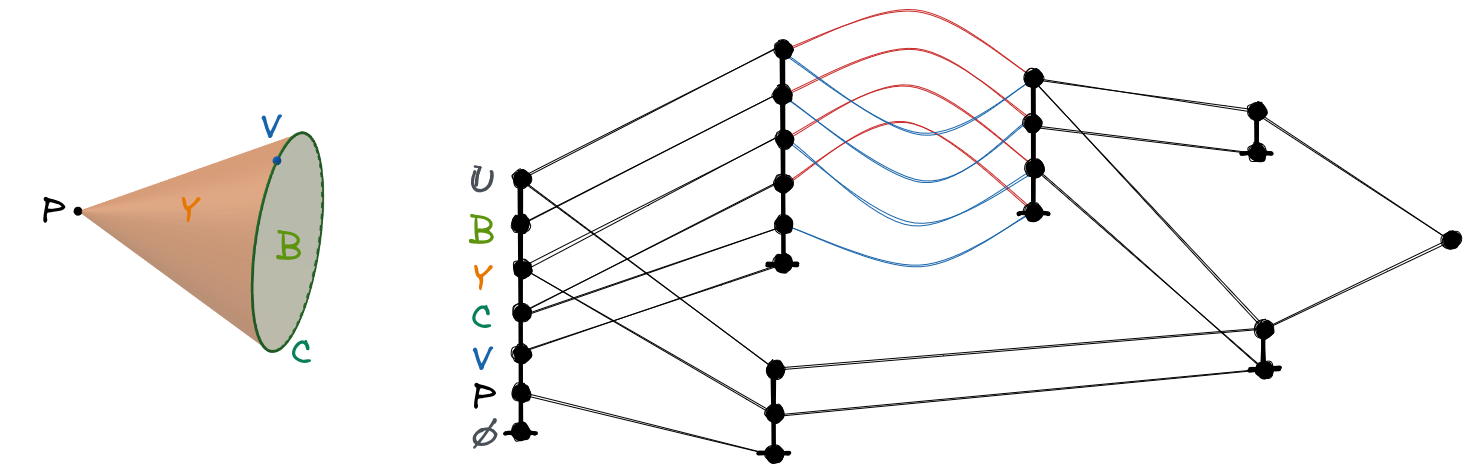}
\caption{The utility pole diagram (right) of a cone (left). Poles
indicate upper categories, points on each pole indicate objects of
corresponding upper category, and wires between points indicate mappings
between objects.}\label{fig:utility-pole}
}
\end{figure}

One can visualize this model as a graph-like diagram, which we call
\textbf{utility pole diagram}. Draw each node of graph as a vertical
line, and put symbols as points on it, which indicate objects of an
upper category. One should put the initial object at the bottom. Draw
the edge of graph as serial wires connecting points on the nodes, which
indicate mappings between objects. The direction of edges should be to
the right. For example, the utility pole diagram of a cone, which is
composed by seven facets, can be drawn as seven poles connected by
wires, see Figure \ref{fig:utility-pole}.

\begin{figure}
\hypertarget{fig:support}{%
\centering
\includegraphics[width=0.3\textwidth,height=\textheight]{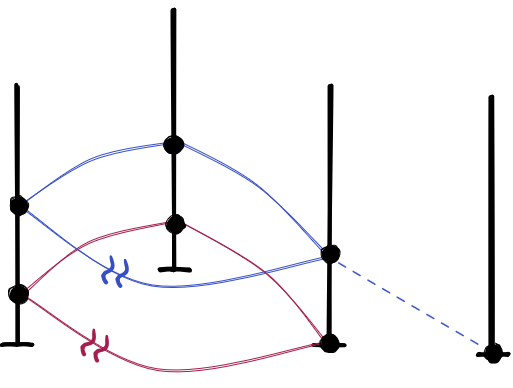}
\caption{Supportivity of wires. The wires (blue lines) are supported by
the base wires (red lines). The broken lines (drawn as \(\approx\))
means there may be multiple poles in between.}\label{fig:support}
}
\end{figure}

We use the curvatures and colors to distinguish edges with the same
parent and child nodes. The mappings of composite functors can be
obtained by following wires through a certain paths. A path of this
graph can be indicated by a string of wires ended at the bottom of a
pole, which are called \textbf{base wires}. The left endpoint of the
base wires of a path indicates the identity of the corresponding
functor: if the base wires of two paths start at the same point, they
should end at the same node on the right, and the mappings of paths
should be the same. That means the circle formed by the base wires of
two paths can be lifted up to any right endpoint. This property is
called \textbf{supportivity}; the wires of a path are said to be
supported by the base wires (see Figure \ref{fig:support}).

\hypertarget{implementation}{%
\subsection{Implementation}\label{implementation}}

In this model, a bounded acyclic category can be fully described by all
nodes and weighted edges, where nodes represent upper categories, and
weighted edges represent downward functors. However, a node cannot fully
describe an upper cateogry, instead, all forwardly reachable nodes
should be included, which is called an upper subgraph. An upper subgraph
also represents a bounded acyclic category, so this model can be encoded
as a recursive data structure: define a data type \texttt{BAC}, which is
composed by a root node and some child BACs, where the root node is
connected to each root node of the child BACs by a type \texttt{Edge}.
Written in a pseudocode, it is defined as

\begin{verbatim}
define type BAC {
    field node: Node,
    field children: List of (Edge, Pointer to BAC),
}

define type Node {
    field symbols: Set of Symbol,
}

define type Edge {
    field dictionary: Map from Symbol to Symbol,
}
\end{verbatim}

The field \texttt{children} is a list of tuple of an \texttt{Edge} and a
pointer to the subgraph, where the edge is from the field \texttt{node}
of this graph to the field \texttt{node} of the subgraph. Under this
definition, each BAC represents an acyclic directed multigraph, which is
defined by unioning smaller graphs recursively. This structure is
similar to the recursive definition of a tree, except that the child
BACs are shared. But in fact, it really is almost a tree, as explained
below.

BACs cannot be mutable referenced. Consider a BAC \texttt{bac} with two
different child BACs \texttt{subbac1} and \texttt{subbac2}, whose
identities are determined by above process, and assume that there is a
parent BAC \texttt{superbac} such that these two child BACs correspond
to the same upper category with respect to \texttt{superbac}. If BACs
are mutable referenced, the in-place mutation of the child BAC
\texttt{subbac1} with respect to \texttt{bac} will ruin the structure of
the parent BAC \texttt{superbac}, since \texttt{subbac1} and
\texttt{subbac2} are now different, which is fine for \texttt{bac} but
not for \texttt{superbac}. If BACs are immutable referenced, there is no
such problem since child BACs can only be modified by replacement.

References to BAC should be implicit. The same subgraphs don't represent
the same upper category, but the same upper category should be
represented by the same instance of BAC. To determine which BAC
indicates which upper category with respect to a given node, one should
determine the identity of corresponding downward functors. If there are
two paths their functors map an initial object to the same object, these
two functors should be the same, since the target object (the initial
morphism of the target object) represents the identity of the functor.
Because the identity of upper category is determined by downward
functors, addresses between child BACs are meaningless to corresponding
bounded acyclic category. References to nodes should therefore be
immutable and implicit, and such definition of BAC is no different than
the recursive definition of tree except for data manipulation.

There are two ways to implement this data structure according to
mutability. A bounded acyclic category can be implemented as a normal
tree structure, where child BACs are implicitly shared. ``Implicitly
shared'' mean that users should not know who shared this BAC with who,
since it is meaningless. In this implementation, operations should be
carefully designed so that two paths representing identical functor
still point to the same BAC. As a consequence, it is preferred to use
immutable data structure and functional programming technique. It can
also be implemented as a directed acyclic graph, where child BACs are
explicitly shared but encapsulated like implicitly shared BACs. In this
implementation, operations should be carefully designed so that BACs
will be copied on write when there are multiple paths which represent
different functors pointing to this BAC. In this article, the former
implementation will be used.

Due to the immutability nature of the first implementation, it is
suitable to use Haskell to show how it works. The purpose of the code in
this work is just a proof of concept, so performance and debuggability
are not concerned. The source code will be placed in the repository
\url{https://github.com/worldmaker18349276/bac}. This data structure is
implemented as a weighted tree with symbol maps as weights:

\begin{Shaded}
\begin{Highlighting}[]
\KeywordTok{newtype} \DataTypeTok{Tree}\NormalTok{ e }\OtherTok{=} \DataTypeTok{Tree}\NormalTok{ (}\DataTypeTok{Map}\NormalTok{ e (}\DataTypeTok{Tree}\NormalTok{ e)) }\KeywordTok{deriving}\NormalTok{ (}\DataTypeTok{Eq}\NormalTok{, }\DataTypeTok{Ord}\NormalTok{, }\DataTypeTok{Show}\NormalTok{)}

\KeywordTok{type} \DataTypeTok{BAC} \OtherTok{=} \DataTypeTok{Tree} \DataTypeTok{Dict}

\KeywordTok{type} \DataTypeTok{Dict} \OtherTok{=} \DataTypeTok{Map} \DataTypeTok{Symbol} \DataTypeTok{Symbol}

\KeywordTok{type} \DataTypeTok{Symbol} \OtherTok{=} \DataTypeTok{Natural}
\end{Highlighting}
\end{Shaded}

The instances of BAC, called nodes, represent bounded acyclic
categories, whose objects are marked by type
\VERB|\DataTypeTok{Symbol}|. Some nodes are implicitly shared, but due
to referential transparency, users cannot know the physical identity of
the node, so one can only say that they are exactly the same. Symbols on
a node are not included in the data structure since it can be determined
by the weights of edges. The symbol list on a node can be obtained by
function \texttt{symbols}, where there must be a special symbol
\texttt{base} representing the initial object of the category. There is
no symbol representing the terminal object since it doesn't provide any
information. Also, in this data structure dealing with non-terminal
morphisms and terminal morphisms are very different.

A type \texttt{Arrow} is defined as a structure of a dictionary and the
target BAC, representing a \emph{local embedding} between categories:

\begin{Shaded}
\begin{Highlighting}[]
\KeywordTok{data} \DataTypeTok{Arrow} \OtherTok{=} \DataTypeTok{Arrow}\NormalTok{ \{}\OtherTok{dict ::} \DataTypeTok{Dict}\NormalTok{,}\OtherTok{ target ::} \DataTypeTok{BAC}\NormalTok{\} }\KeywordTok{deriving}\NormalTok{ (}\DataTypeTok{Eq}\NormalTok{, }\DataTypeTok{Ord}\NormalTok{, }\DataTypeTok{Show}\NormalTok{)}

\OtherTok{edges ::} \DataTypeTok{BAC} \OtherTok{{-}\textgreater{}}\NormalTok{ [}\DataTypeTok{Arrow}\NormalTok{]}
\NormalTok{edges (}\DataTypeTok{Tree}\NormalTok{ m) }\OtherTok{=} \FunctionTok{fmap}\NormalTok{ (}\FunctionTok{uncurry} \DataTypeTok{Arrow}\NormalTok{) (Map.toList m)}
\end{Highlighting}
\end{Shaded}

The edges of a node can be obtained by function \texttt{edges}, which
are represented as type \VERB|\DataTypeTok{Arrow}|. The weight of an
edge, called a \textbf{dictionary}, is a non-empty mapping from the
symbols on child node to the symbols on the parent node, representing
the mapping of the objects. It is a non-empty map because there must be
a base symbol as a key. In a node, dictionaries of its edges must cover
all symbols except the base symbol, so the valid symbols can be
determined by weights of outgoing edges.

A \textbf{path} of the tree, which is a sequence of connected edges,
also has a derived dictionary, and it can be obtained by concatenating
all dictionaries along the path by \texttt{cat}. A path from a node to
itself by following no edge is called a \textbf{null path}. A null path
possesses a trivial dictionary, which is the identity mapping of symbols
of the node. A null path is said to be an improper path. Dictionaries of
paths together with the target nodes represent downward functors between
bounded acyclic categories. It is called an \textbf{arrow} in the
program, which can also be represented by a type
\VERB|\DataTypeTok{Arrow}|.

An arrow represents a downward functor from the category of the child
node to the category of the parent node, where the mapping of objects is
represented by the dictionary of an arrow, and the mapping of morphisms
can be determined by further analysis of the structure of the child
node. As a downward functor, the field \texttt{target} represents an
upper category constructed under the category of the parent node. An
arrow can also represent an initial morphism in the category of the
parent node, where the field \texttt{target} represents the target
object of this morphism. There is an one-to-one correspondence between
objects and initial morphisms, so it also represents an object in the
category, in which the symbol that refers to the object is the symbol to
which the dictionary maps the base symbol, and all other values of the
dictionary represent all descendant objects of this object. Two arrows
starting at the same node should correspond to the same functor if their
dictionaries map the base symbol to the same symbol. In this case, the
target nodes should be implicitly shared or exactly the same. Note that
it's not true the other way around; two arrows that have the same target
node doesn't mean they correspond to the same functor.

In program, arrows are more useful than symbols because they contain
more information. Some algebraic operations can be performed on arrows
by utilizing these information. The arrow of a null path, which
represents the identity functor, can be obtained by \texttt{root}. The
arrow of a path can be obtained by joining all edges along the path,
which is implemented as a function \texttt{join}. One can explore all
upper categories under a given category using these functions, which is
equivalent to explore all objects of this category. Reversely, some
arrows start at the same node are divisible, and the result can be
obtained by function \texttt{divide}: the result of division between the
divisor \texttt{arr12} and the dividend \texttt{arr13} is the set of
arrows \texttt{arr23} such that
\VERB|\NormalTok{join arr12 arr23 }\OtherTok{=}\NormalTok{ arr13}|. As a
node of a tree, an arrow knows whether a symbol refers to its
descendant, which is implemented as a function \texttt{locate}. The
upper category specified by a given symbol can be obtained by tracing a
symbol with an arrow, which is implemented as \texttt{arrow}. Reversely,
the symbol specifying a given arrow can be obtained with the function
\texttt{symbol}, which also indicates the identity of the given arrow.

In a category represented by a node, an object can be specified by a
symbol, and a morphism can be specified by a tuple of symbols: the
source object of the morphism is represented by the first symbol, and it
should be a valid symbol in this node; the target object of the morphism
is represented by the second symbol, and it should be a valid symbol in
the node referenced by the first symbol. The object specified by a
symbol can also be represented by an arrow, which can be obtained by
\texttt{arrow}. Also, the morphism specified by two symbols can be
represented by a tuple of connected arrows, which corresponds to two
connected downward functors. The conversion between the two
representations are implemented as the functions \texttt{arrow2} and
\texttt{symbol2}. In another viewpoint, a symbol indicates an
\(1\)-chain of the host category, and a tuple of symbols indicates a
\(2\)-chain, moreover, a sequence of \(n\) symbols indicates a
\(n\)-chain.

Conclude above descriptions, there are three laws for BAC:

\begin{enumerate}
\def\labelenumi{\arabic{enumi}.}
\tightlist
\item
  \emph{totality}: the dictionary of an edge should be a mapping from
  all valid symbols on the child node to valid symbols on the parent
  node.
\item
  \emph{surjectivity}: all valid symbols should be covered by the
  dictionaries of outgoing edges, except the base symbol.
\item
  \emph{supportivity}: if dictionaries of given two paths with the same
  starting node map the base symbol to the same symbol, then they should
  have the same dictionary and target node. Note that null paths also
  count.
\end{enumerate}

A valid bounded acyclic category should satisfy these laws, and they
immediately derive two laws:

\begin{enumerate}
\def\labelenumi{\arabic{enumi}.}
\setcounter{enumi}{3}
\tightlist
\item
  The dictionary of a proper path cannot maps a symbol to the base
  symbol, otherwise the target node should be the starting node.
\item
  The dictionary of a proper path should map the base symbol to a unique
  symbol compared with other values in this dictionary, otherwise the
  target node will have a descendant as itself.
\end{enumerate}

The essence of recursive types is folding. To traverse objects of a
bounded acyclic category, or equivalently, to traverse upper categories
of a bound acyclic category, all descendant nodes should be visited via
arrows, which is implemented as the function \texttt{fold}. While
folding, identical nodes (from the root's perspective) should only be
visited once, which is important for non-purely functional programming
language (support for side effects without explicit types), otherwise it
is possible to obtain inconsistent results with identical inputs. Users
may more commonly fold data only under a certain symbol (visit only
ancestor nodes of the node referenced by this symbol). In this
situation, the relative location of a node should be marked. It is
defined as the function \texttt{foldUnder}, which only visit the node
that can reach to a given symbol. The fold method can also be used to
modify data structures. Convenient specialized functions \texttt{modify}
and \texttt{modifyUnder} handle how edges should be modified and put
back into nodes. This allows us to edit data in the form:

\begin{Shaded}
\begin{Highlighting}[]
\OtherTok{dosomething ::} \DataTypeTok{BAC} \OtherTok{{-}\textgreater{}} \DataTypeTok{Maybe} \DataTypeTok{BAC}
\NormalTok{dosomething node }\OtherTok{=} \KeywordTok{do}
  \CommentTok{{-}{-} check if this operation is valid}
\NormalTok{  guard }\OperatorTok{$} \OperatorTok{...}

  \CommentTok{{-}{-} edit the boundary node}
  \KeywordTok{let}\NormalTok{ res0 }\OtherTok{=} \OperatorTok{...}

  \CommentTok{{-}{-} edit edges under the boundary node}
\NormalTok{  fromReachable res0 }\OperatorTok{$}\NormalTok{ node }\OperatorTok{\&}\NormalTok{ modifyUnder src \textbackslash{}(curr, edge) }\OtherTok{{-}\textgreater{}}\NormalTok{ \textbackslash{}}\KeywordTok{case}
    \CommentTok{{-}{-} if the edge points to an outer node}
    \DataTypeTok{AtOuter} \OtherTok{{-}\textgreater{}} \OperatorTok{...}
    \CommentTok{{-}{-} if the edge points to the boundary node}
    \DataTypeTok{AtBoundary} \OtherTok{{-}\textgreater{}} \OperatorTok{...}
    \CommentTok{{-}{-} if the edge points to an inner node, which has been modified to \textasciigrave{}res\textasciigrave{}}
    \DataTypeTok{AtInner}\NormalTok{ res }\OtherTok{{-}\textgreater{}} \OperatorTok{...}
\end{Highlighting}
\end{Shaded}

With those utility functions, it should be easy to manipulate BAC. Below
we will discuss how to edit a BAC from geometrical, categorical and
programming perspectives.

\hypertarget{fundamental-operations}{%
\section{Fundamental Operations}\label{fundamental-operations}}

The data structure BAC can be used to encode the incidence structures of
geometric objects. In the BAC, symbols on the root node represents
geometric objects, and arrows represent incidence relations between
geometric objects. This implementation does not encode any meaning of
arrows, only the relation and composition between arrows. Users should
associate arrows and incidence relations outside of this data structure.
Below we will focus on algebraic properties of incidence relations.

In order to manipulate geometric objects in the form of bounded acyclic
categories, all geometric operations need to be rewritten. To simplify
the whole workflow, one need to develop a set of basic operations, such
as boolean operations. Even for the intersection operation, the actions
in categorical level are complicated. For example, to put two facets
together, not only the relationship between two facets is needed, the
relationship between their subfacets, and subfacets of subfacets are
also needed. So we need to further decompose boolean operations into
more fundamental operations on category. These operations may not make
sense on geometrical perspective, but they are critical on categorical
level.

Let's discuss how to intersect two balls: we just need to slice a ball
by a sphere, then remove the unwanted part. Before start slicing, how
did the ball come into our world? Moreover, how did the world began? We
need an simplest polytope as our starting point to construct all kind of
polytopes. Such polytope is called \textbf{``nullitope''}, which is a
geometric object of nothing. Then we need a method to add a ball into
this polytope. This operation is called \textbf{``introduce''}. Now we
want to slice this ball by a sphere, which should result in a UFO shape.
But before that, we need to compute the intersection of their surface,
which is a circle. We then can use ``introduce'' to bring the circle
into our world, and claim that this circle is covered by those two
spheres. This operation is called \textbf{``incident''}. Now we have two
spheres with a circle on it, but wait, doesn't it is disconnected? It
should be separated as a cap and a cup, so we need an operation called
\textbf{``disconnect''}. Now we have two caps to build our UFO shell,
but how to get rid of those two cup? To do that, we need
\textbf{``remove''} to clean up them.

Ok, there are enough tools to intersect two balls, let's describe it
step-by-step.

\begin{enumerate}
\def\labelenumi{\arabic{enumi}.}
\tightlist
\item
  Prepare an empty space by ``nullitope''.
\item
  Prepare a ball.

  \begin{itemize}
  \tightlist
  \item
    Create an infinite 3D space by ``introduce''.
  \item
    Create a sphere with radius \texttt{2} and centered at
    \texttt{(0,0,1)} by ``introduce'', labeled as \texttt{S}.
  \item
    Claim that the sphere \texttt{S} is covered by this 3D space using
    ``incident''.
  \item
    Separate inner space and outer space by sphere \texttt{S} as two
    disconnected components using ``disconnect''.
  \item
    Remove the outer space by ``remove'', and label the remaining one as
    \texttt{V}.
  \end{itemize}
\item
  Create a sphere with radius \texttt{2} and centered at
  \texttt{(0,0,-1)} by ``introduce'', labeled as
  \texttt{S\textquotesingle{}}.
\item
  Cut the ball \texttt{V} by this sphere \texttt{S\textquotesingle{}}.

  \begin{itemize}
  \tightlist
  \item
    Compute the intersection between shell \texttt{S} and
    \texttt{S\textquotesingle{}}, which is a circle.
  \item
    Bring this circle into the world by ``introduce'', labeled as
    \texttt{C}.
  \item
    Claim that the circle \texttt{C} is covered by spheres \texttt{S}
    and \texttt{S\textquotesingle{}} using ``incident''.
  \item
    Separate spheres \texttt{S} and \texttt{S\textquotesingle{}} into
    cups and caps by ``disconnect'', labeled as \texttt{S1},
    \texttt{S2}, \texttt{S1\textquotesingle{}},
    \texttt{S2\textquotesingle{}} separately.
  \item
    Remove the cup \texttt{S1\textquotesingle{}} by ``remove''.
  \item
    Claim that the cap \texttt{S2\textquotesingle{}} is covered by
    volume \texttt{V} using ``incident''.
  \item
    Separate volume \texttt{V} into two part using ``disconnect'',
    labeled as \texttt{V1}, \texttt{V2}.
  \end{itemize}
\item
  Remove unwanted part.

  \begin{itemize}
  \tightlist
  \item
    Remove outer volume \texttt{V2} by ``remove''.
  \item
    Remove cup \texttt{S1} by ``remove''.
  \end{itemize}
\end{enumerate}

Those operations have corresponding categorical meanings, which provide
another perspective to investigate fundamental operations. Now let's
discuss how to define those operations.

\hypertarget{emptysingleton}{%
\subsection{Empty/Singleton}\label{emptysingleton}}

\textbf{``Nullitope''} is the simplest polytope, which is a geometric
object of nothing. It is just like an empty workspace after launching
GeoGebra 3D calculator. In categorical perspective, it has only two
objects: the initial object and the terminal object. This category is
called \textbf{empty bounded acyclic category}. The empty bounded
acyclic category has only one non-degenerated morphism chain, and it is
locally-embedded into any nondecomposable morphism of a bounded acyclic
category. In program, it is implemented as a function \texttt{empty},
which is a node without children.

\textbf{``Singleton''} is the second simplest polytope, which represents
a geometric object without boundary. It is just like an primitive shape
in the GeoGebra 3D calculator, such as a sphere, an infinite plane or a
circle. In categorical perspective, it has only one proper object. This
category is called \textbf{singleton bounded acyclic category}. In
program, it is implemented as a function \texttt{singleton}, which is a
node with only one child.

\hypertarget{merge-categories}{%
\subsection{Merge Categories}\label{merge-categories}}

\textbf{``Introduce''} bring a facet into the universe, just like we can
simply drop a sphere, a cube or a cylinder into the workspace. For
example, the process of intersecting two surfaces is done by adding
intersected line and claiming the incidence relations between them.
After adding intersected line, the computer doesn't know the relation
between this line and those surfaces. This doesn't break the categorical
structure, so there is no need to add incidence relations at the same
time. In categorical perspective, it merges two categories into one,
where the initial object and the terminal object are merged, and others
proper objects are considered disjoint. In program, it corresponds to
merging trees at the root. All symbols in the root nodes are unioned
except the base symbols, which will be merged into one base symbol. It
is implemented as a function \texttt{mergeRootNodes} with a parameter
\VERB|\OtherTok{nodes ::}\NormalTok{ [}\DataTypeTok{BAC}\NormalTok{]}|,
representing the categories to merge.

\hypertarget{remove-a-terminal-morphism}{%
\subsection{Remove a Terminal
Morphism}\label{remove-a-terminal-morphism}}

\begin{figure}
\hypertarget{fig:remove-node}{%
\centering
\includegraphics[width=0.3\textwidth,height=\textheight]{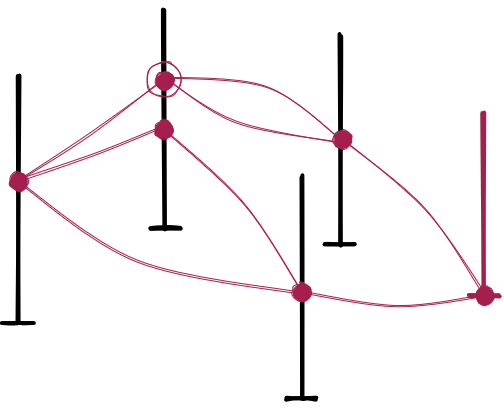}
\caption{Remove a leaf node. Removing a node (red pole) also removes all
points and wires connected to it (red points and red lines). In this
example, there is a point (circled point) that connects to the removed
node by multiple paths, exhibiting a non-trivial structure. Removing the
node will drop such information.}\label{fig:remove-node}
}
\end{figure}

\textbf{``Remove''} removes a facet out of the universe. The subfacets
of removed facet will not be removed. For example, if a square is
removed, its borders are not removed along with it. The removed facet
cannot be covered by other facets (not a subfacet of anther facet). In
categorical perspective, it removes a nondecomposable terminal morphism
of the given object, and as a consequence, the morphisms pointing to
this object are also removed. In program, this process corresponds to
removing a leaf node specified by a symbol. Since it is a leaf node,
there is only one symbol (the base symbol) on this node. All wires that
connect to the base point of removed node should be removed, so all
passing points will also be removed. Information related to those wires
will be lost (see Figure \ref{fig:remove-node}). It is implemented as a
function \texttt{removeLeafNode} with a parameter
\VERB|\OtherTok{tgt ::} \DataTypeTok{Symbol}|, indicating the object to
remove. The object specified by the symbol \texttt{tgt} should have
nondecomposable terminal morphism.

\hypertarget{remove-a-non-terminal-morphism}{%
\subsection{Remove a Non-Terminal
Morphism}\label{remove-a-non-terminal-morphism}}

In categorical perspective, ``remove'' removes a nondecomposable
terminal morphism. It is natural to generalize it to any nondecomposable
morphism. To remove a nondecomposable morphism, all composition rules in
the multiplication table related to this morphism will also be removed.
For example, to remove morphism \(\phi\), the rule
\(\phi \circ \phi' = \psi\) for any possible \(\phi'\) and \(\psi\) will
also be removed.

However, removing \emph{decomposable} morphism is not simple; there are
multiple ways to remove a decomposable morphism consistently. For
example, to remove morphism \(\phi\) while there is a composition rule
\(\phi = \phi_1 \circ \phi_2\), one of morphism \(\phi_1\) or \(\phi_2\)
should also be removed: one can remove \(\phi_1\) then \(\phi\), remove
\(\phi_2\) then \(\phi\), or remove both \(\phi_1\) and \(\phi_2\) then
\(\phi\). Two coherent choices are: removing all prefix ones, or
removing all suffix ones. This process can be decomposed into multiple
steps of removing nondecomposable morphisms. Here only nondecomposable
morphisms are concerned.

In geometrical perspective, it is called \textbf{``unincident''} because
it corresponds to removing an incidence relation. Removing an initial
morphism leads to the target object being removed, which corresponds to
removing a subfacet of some facets, and such subfacet should be
boundaryless. It is the same as ``remove'' if it is not a subfacet of
any facet. Removing a non-initial morphism can be understood as removing
a subfacet of some facets from a vertex figure, which results in
``unincident'' some facets. Removing a subfacet can lead to some ill
geometric shapes. For example, it is nonsense to remove the boundary
circle of a disk. If the subfacet to be removed has a 2-dimension
difference from the facet, such as a point covered by a plane, it is
usually fine, otherwise it should be carefully dealt with. For example,
a point on the circle can be safely removed, but removing the endpoint
of a line segment leads to an ill geometric object.

\begin{figure}
\hypertarget{fig:remove-symbol}{%
\centering
\includegraphics[width=0.5\textwidth,height=\textheight]{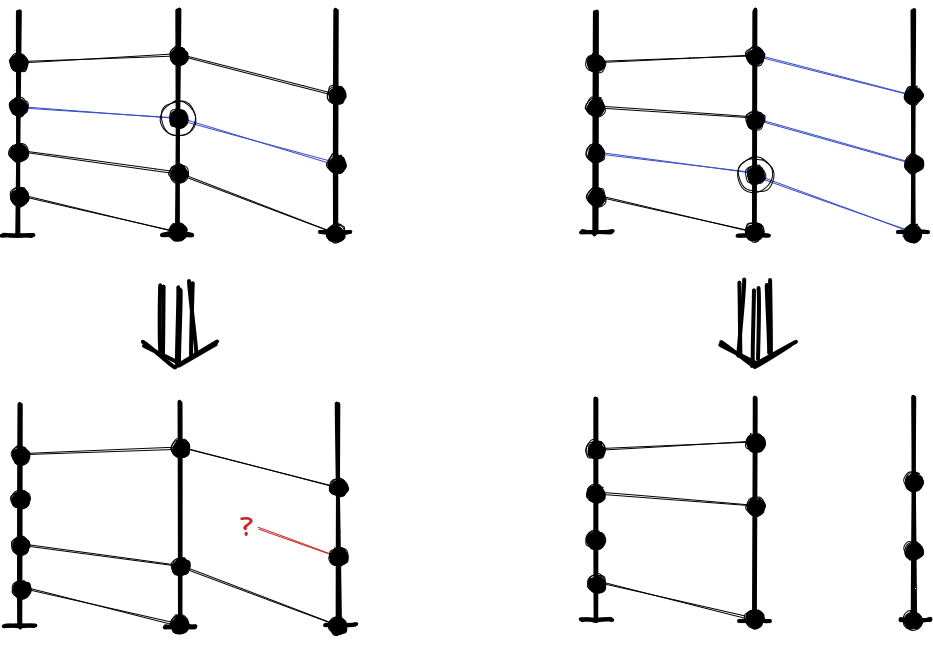}
\caption{Remove a symbol (circled point) on a node. Left one shows that
the removed symbol cannot be decomposable: there is an outgoing wire
connected to a non-base point, and such wire will be dangling (red line)
when the point is removed. Right one shows that removing a
nondecomposable symbol will lead to removing the entire edge (blue
lines), since the base wire is removed.}\label{fig:remove-symbol}
}
\end{figure}

\begin{figure}
\hypertarget{fig:alternative-paths}{%
\centering
\includegraphics[width=0.5\textwidth,height=\textheight]{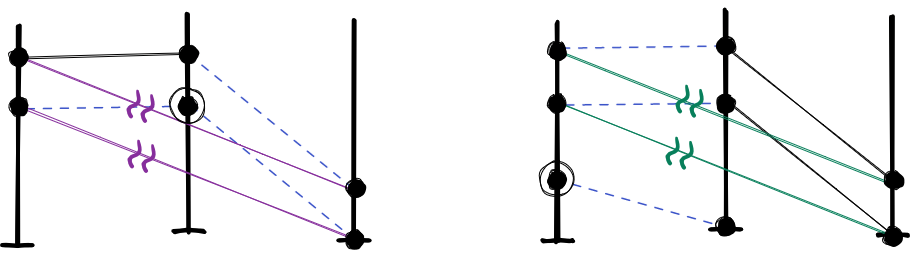}
\caption{Alternative paths for the removed/added wires (blue dashed
lines). Left one shows alternative wires (purple lines) ending at the
target node of the removed/added edge. Right one shows alternative wires
(green lines) starting at the source node of the removed/added edge. The
removed/added wires are related by supportivity, which form parallel
triangles.}\label{fig:alternative-paths}
}
\end{figure}

In program, this process removes a symbol on a node, and keep all other
unrelated wires unchanged. Where ``keep all other unrelated wires
unchanged'' means that only the outgoing and incoming edges of this node
will be modified. Consider a base wire passing through but not ending at
this point, the starting point of this wire should not change after this
process. It is a minimal operation, since only a small part of this data
structure is changed. By removing a symbol on a node, the adjacent wires
will also be removed. It's fine to remove incoming wires, but invalid
for removing outgoing wires, except when outgoing wires is connected to
the base point, since now the entire outgoing edge should be removed
(see Figure \ref{fig:remove-symbol}). Symbols connected only to the base
points in the right are said to be \textbf{nondecomposable}, which
represent nondecomposable initial morphisms. Above discussion shows only
nondecomposable symbols can be removed. All wires that crossed the
removed wires should find alternative paths. For all parent nodes of the
source node of the removed edge, one only need to check base wires
connected to the target node of the removed edge (see left one of Figure
\ref{fig:alternative-paths}). Similarly, for all child nodes of the
target node of the removed edge, one only need to check base wires
connected to this child node (see right one of Figure
\ref{fig:alternative-paths}). It is equivalent to: surjectivity of
dictionaries should not be violated after removing this edge.

It is implemented as a function \texttt{removeNDSymbol} with a parameter
\VERB|\NormalTok{(src, tgt)}\OtherTok{ ::}\NormalTok{ (}\DataTypeTok{Symbol}\NormalTok{, }\DataTypeTok{Symbol}\NormalTok{)}|,
indicating the morphism to remove, which should be nondecomposable. The
decomposability of a morphism can be checked by the function
\texttt{nondecomposable}.

\hypertarget{add-a-non-terminal-morphism}{%
\subsection{Add a Non-Terminal
Morphism}\label{add-a-non-terminal-morphism}}

\textbf{``Incident''} is a method to claim the incidence relations
between two facets. This method is the reverse of ``unincident''. It is
just like attach a point to a surface in the GeoGebra 3D calculator, but
here we don't change the position of the point. Notice that ``a line
segment is covered by a plane'' implies ``the endpoints of this line
segment are also covered by this plane''. To claim A is covered by B,
all objects covered by A should be already covered by B. In this
example, we need to claim ``the endpoints are covered by this plane''
first. But to do that, all subfacets of endpoints, which is the null
face, should be covered by this plane, and this is already true. In the
opposite, ``a point is covered by an edge of a square'' implies ``this
point is also covered by this square''. To claim A is covered by B, all
objects that cover B should already cover A. In this example, we need to
claim ``this point is covered by this square'' first. But to do that,
all superfacet of this square, which is the universe, should cover this
point, and this is already true.

\begin{figure}
\hypertarget{fig:dried-persimmon}{%
\centering
\includegraphics[width=0.6\textwidth,height=\textheight]{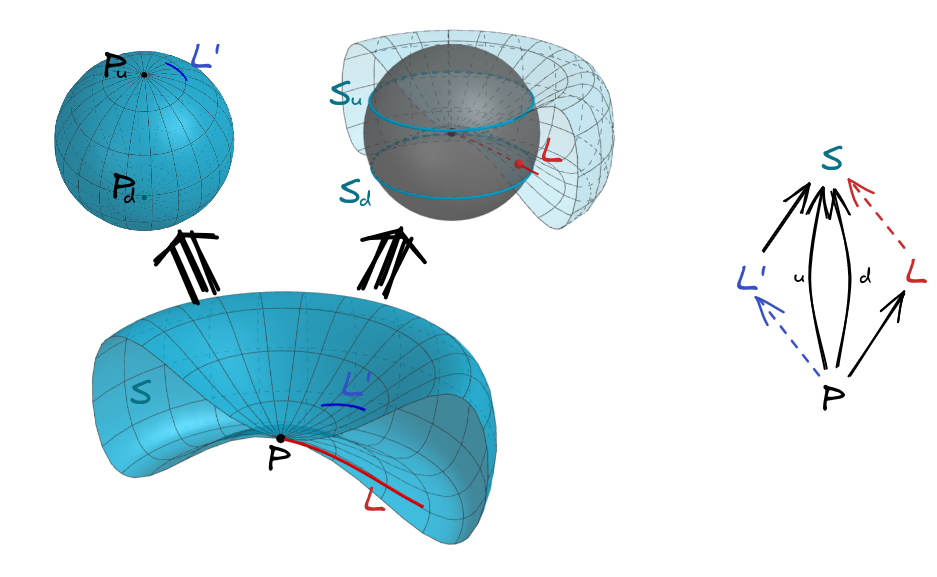}
\caption{Ambiguity in claiming incidence relations in the dried
persimmon shape (left), which are equivalent to adding morphisms between
two objects (right). To claim \texttt{L} is covered by \texttt{S} (to
add red dashed line in the right diagram), the vertex figure should be
considered (right sphere). To claim \texttt{P} is covered by
\texttt{L\textquotesingle{}} (to add blue dashed line in the right
diagram), the face figure should be considered (left
sphere).}\label{fig:dried-persimmon}
}
\end{figure}

In some case, it is not enough to claim incidence relations by just
saying ``A is covered by B''. For example, the dried persimmon shape
described by equation \(z^2 = (1-x^2-y^2)(x^2+y^2)^2\) is a sphere but
north and south poles are pinched together, so the center point is
covered by this surface in two directions. Consider a curved line
segment starting at the center point, to claim ``this segment is covered
by this dried persimmon shape'', it is also necessary to specify ``which
side of this shape''. In the vertex figure of the center point, the
problem becomes that ``this point is covered by which one of two
circles'' (see right sphere of Figure \ref{fig:dried-persimmon}). In the
opposite, consider a curved line segment on this dried persimmon shape,
to claim ``this segment covers the center point'', it is also necessary
to specify ``which side of this shape''. In the face figure of this
shape, the problem becomes that ``this segment covers which one of two
points'' (see left sphere of Figure \ref{fig:dried-persimmon}).

In categorical perspective, this process just add a non-terminal
morphism into this category between two given objects. Assume we are
adding a morphism \(\phi : F \rightarrow S\), and if there exists
another morphism \(\phi' : P \rightarrow F\) but no morphism such that
\(? : P \rightarrow S\), we will also need to add another morphism
\(\phi'' : P \rightarrow S\) such that \(\phi'' = \phi \circ \phi'\),
otherwise it will break the closure property of category; it is the
first point we discussed above. Similarly, if there exists another
morphism \(\phi' : S \rightarrow V\) but no morphism such that
\(? : F \rightarrow V\), we will also need to add another morphism
\(\phi'' : F \rightarrow V\) such that \(\phi'' = \phi' \circ \phi\); it
is the second point we discussed above. So we should add those morphisms
first, and it is possible by recursively repeating it down/up to the
initial/terminal object. After confirming that closure property can be
retained, we need to expand the multiplication table of morphisms. We
only need to provide how to compose nondecomposable morphisms, since
other cases can always be decomposed into this case. Even though this
process can be automatically done in the most case, but for non-trivial
case it becomes necessary.

The choices of new composition rules should agree with the existing
equivalence relations: if we add a morphism \(\phi : S \rightarrow F\)
and claim relations \(\phi_1' = \phi \circ \phi_1\) and
\(\phi_2 = \phi_2' \circ \phi\), where \(\phi_1 : P \rightarrow S\),
\(\phi_2 : S \rightarrow V\), \(\phi_1' : P \rightarrow F\),
\(\phi_2' : F \rightarrow V\), the equivalence relation
\(\phi_2 \circ \phi_1 = \phi_2' \circ \phi_1'\) should already exist.
Also, if we claim that \(\phi_1' = \phi \circ \phi_1\) and
\(\phi_2' = \phi \circ \phi_2\), where \(\phi_1 : P_1 \rightarrow S\)
and \(\phi_2 : P_2 \rightarrow S\), and there exists two morphisms
\(\psi_1 : V \rightarrow P_1\) and \(\psi_2 : V \rightarrow P_2\), such
that \(\phi_1 \circ \psi_1 = \phi_2 \circ \psi_2\), then the equivalence
relation \(\phi_1' \circ \psi_1 = \phi_2' \circ \psi_2\) should already
exists. A dual version should also hold. They can be concluded into
three laws:

\begin{enumerate}
\def\labelenumi{\arabic{enumi}.}
\tightlist
\item
  \(\phi_R' = \phi \circ' \phi_R \land \phi_L' = \phi_L \circ' \phi \implies \phi_L' \circ \phi_R = \phi_L \circ \phi_R'\).
\item
  \(\phi_{R1}' = \phi \circ' \phi_{R1} \land \phi_{R2}' = \phi \circ' \phi_{R2} \land \phi_{R1} \circ \zeta_{R1} = \phi_{R2} \circ \zeta_{R2} \implies \phi_{R1}' \circ \zeta_{R1} = \phi_{R2}' \circ \zeta_{R2}\).
\item
  \(\phi_{L1}' = \phi_{L1} \circ' \phi \land \phi_{L2}' = \phi_{L2} \circ' \phi \land \zeta_{L1} \circ \phi_{L1} = \zeta_{L2} \circ \phi_{L2} \implies \zeta_{L1} \circ \phi_{L1}' = \zeta_{L2} \circ \phi_{L2}'\).
\end{enumerate}

Where \(\circ'\) is the composition related to \(\phi\), which decides
how to compose the new morphism with old one:

\begin{enumerate}
\def\labelenumi{\arabic{enumi}.}
\tightlist
\item
  For \(\phi_R : S' \rightarrow S\), \(\phi_R' = \phi \circ' \phi_R\) is
  a morphism \(\phi_R' : S' \rightarrow F\).
\item
  For \(\phi_L : F \rightarrow F'\), \(\phi_L' = \phi_L \circ' \phi\) is
  a morphism \(\phi_L' : S \rightarrow F'\).
\end{enumerate}

One only need to define the composition rule for all nondecomposable
ones of \(\phi_R\) and \(\phi_L\). There are multiple choices of
\(\phi_R' : S' \rightarrow F\) as a result of \(\phi \circ' \phi_R\),
and each choice can be denoted as a tuple \((\phi_R, \phi_R')\) called a
\textbf{coangle}. Similarly, a choice of the composition rule in another
direction can be denoted as a tuple \((\phi_L, \phi_L')\) called an
\textbf{angle}. Angles and coangles form a vertex set, where they are
grouped by \(\phi_R\) and \(\phi_L\). To make a composition rule, one
should choose a vertex for each group.

The third constraint with \(\phi_{L1} = \phi_{L2}\) excludes some
angles. A \textbf{fork} of a morphism \(\phi\) is a pair of distinct
morphisms \(\psi\), \(\psi'\) such that
\(\psi \circ \phi = \psi' \circ \phi\). A angle \((\phi_L, \phi_L')\) is
valid if forks of the morphism \(\phi_L\) are also forks of the morphism
\(\phi_L'\). Similarly, some coangles are excluded by the second
constraint.

The first constraint limits how to choose \(\phi_L'\) and \(\phi_R'\)
for each pair of \(\phi_L\) and \(\phi_R\). For a valid choice of pair
\((\phi_L', \phi_R')\), we draw an edge between the angle
\((\phi_L, \phi_L')\) and the coangle \((\phi_R, \phi_R')\). Such
construction makes the induced subgraph of vertex groups \(\phi_L\) and
\(\phi_R\) forms a biclique cover (disjoint union of complete bipartite
graphs).

The second constraint also limits how to choose \(\phi_{R1}'\) and
\(\phi_{R2}'\) for each pair of \(\phi_{R1}\) and \(\phi_{R2}\). Define
\textbf{pseudo-equalizer} between \(\phi_{R1}\) and \(\phi_{R2}\), which
is a pair of morphisms \((\psi, \psi')\) such that
\(\phi_{R1} \circ \psi = \phi_{R2} \circ \psi'\). Two coangles
\((\phi_{R1}, \phi_{R1}')\) and \((\phi_{R2}, \phi_{R2}')\) are
compatible if all pseudo-equalizers between \(\phi_{R1}\) and
\(\phi_{R2}\) are also pseudo-equalizers between \(\phi_{R1}'\) and
\(\phi_{R2}'\). Note that one only need to check all minimal
pseudo-equalizers between \(\phi_{R1}\) and \(\phi_{R2}\). For a valid
choice of pair \((\phi_{R1}', \phi_{R2}')\), we draw an edge between
coangles \((\phi_{R1}, \phi_{R1}')\) and \((\phi_{R2}, \phi_{R2}')\),
and the induced subgraph also forms a biclique cover. The same analysis
can be applied to the third constraint.

Finally, it leads to a graph problem: find a complete subgraph of a
semi-complete multipartite graph by selecting a vertex in each group. A
\textbf{semi-complete multipartite graph} is a multipartite graph where
the induced subgraph of each pair of groups is a biclique cover. In the
practical application of geometry, it is not necessary to actually find
all possible solutions of this problem, since selecting are done by
geometric calculations. This constraint can be used to check if the
geometric calculations are conflicting, or reduce the amount of
calculations by some strategies: one can utilize information theory to
estimate the entropy of selection event for each group, and calculate
the largest one first.

\begin{figure}
\hypertarget{fig:add-symbol}{%
\centering
\includegraphics[width=0.5\textwidth,height=\textheight]{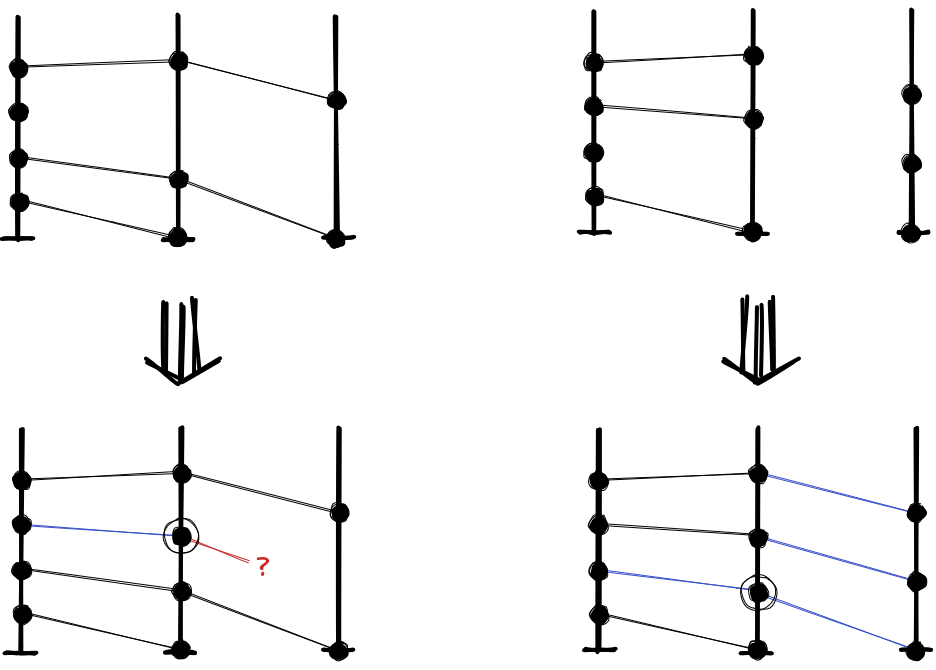}
\caption{Add a symbol (circled point) to a node. Left one shows that the
added symbol should be nondecomposable: there is no way to add a wire
(red line) for an existing edge. Right one shows that adding a
nondecomposable symbol will lead to adding the entire edge (blue
lines).}\label{fig:add-symbol}
}
\end{figure}

In program, this process adds a symbol to a node, and keep all other
unrelated wires unchanged. The wires connected to this point should also
be added. It's fine to add an incoming wire, but it is impossible to add
an outgoing wire to an existing outgoing edge. That means the added
symbol should be nondecomposable, and a new edge should be added such
that the base wire is connected to this point (see Figure
\ref{fig:add-symbol}). The target node of the added edge may be
unreachable from the given node. One should notice that the added edge
shouldn't form a directed loop in the graph.

\begin{figure}
\hypertarget{fig:add-wires-determine}{%
\centering
\includegraphics[width=0.4\textwidth,height=\textheight]{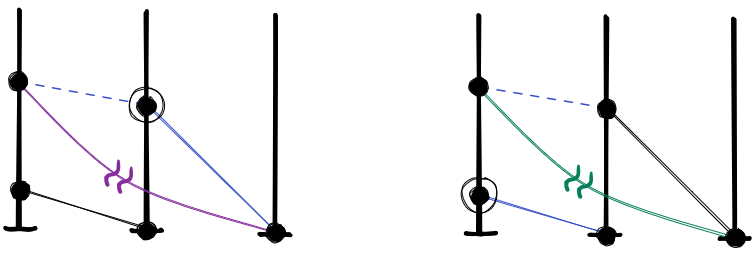}
\caption{Determine added wires (blue dashed line). To determine an added
incoming wire (left diagram), a base wire (purple line) should be
specified. To determine an added outgoing wire (right diagram), a base
wire (green line) should be specified.}\label{fig:add-wires-determine}
}
\end{figure}

\begin{figure}
\hypertarget{fig:add-wires-constraint}{%
\centering
\includegraphics[width=0.2\textwidth,height=\textheight]{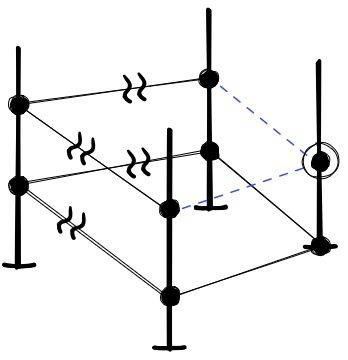}
\caption{Another constraint for adding wires. Each pair of added
incoming wires (blue dashed lines) should be supported by a base
circle.}\label{fig:add-wires-constraint}
}
\end{figure}

Each added wire can be determined by following process: for any pair of
connected edges in which one is the added edge, a base wire of this path
should be specified so that a wire can be added like Figure
\ref{fig:add-wires-determine}. Some wires are determined by
supportivity. Such choice is not arbitrary, supportivity must still hold
after adding wires. For a pair of connected edges in which the second is
the added edge (left one of Figure \ref{fig:alternative-paths}), its
base wire (lower purple line) together with two wires form a triangle,
and all added outgoing wires (blue dashed lines) should be supported by
this triangle. Different triangles must agree with each other.
Similarly, for a pair of connected edges in which the first is the added
edge (right one of Figure \ref{fig:alternative-paths}), its base wire
(lower green line) together with two wires form a triangle, and some
added outgoing wires (blue dashed lines) should be supported by this
triangle. Different triangles should agree with each other. Moreover, if
there is a circle formed by two base wires of the target node (see
Figure \ref{fig:add-wires-constraint}), a pair of corresponding added
wires should be supported by such circle.

The possible choices can be obtained by a helper function
\texttt{findValidCoanglesAngles}, which returns two groups of picklists.
The second group contains picklists of angles, which are used to
determine the outgoing wires (right one of Figure
\ref{fig:add-wires-determine}). The first group contains picklists of
coangles, which are used to determine the incoming wires (left one of
Figure \ref{fig:add-wires-determine}). User should select one of angle
or coangle for each picklist. This does not guarantee a valid choice. A
valid choice of angles and coangles can be checked by functions
\texttt{compatibleAngles}, \texttt{compatibleCoangles} and
\texttt{compatibleCoanglesAngles}.

The process of adding a non-terminal morphism is implemented as a
function \texttt{addNDSymbol} with parameters
\VERB|\NormalTok{src, tgt,}\OtherTok{ sym ::} \DataTypeTok{Symbol}| and
\VERB|\OtherTok{src\_alts ::}\NormalTok{ [}\DataTypeTok{Coangle}\NormalTok{]}|
and
\VERB|\OtherTok{tgt\_alts ::}\NormalTok{ [}\DataTypeTok{Angle}\NormalTok{]}|.
\texttt{src} and \texttt{tgt} indicate source object and target object
of the added morphism, and \texttt{(src,\ sym)} will indicate the added
morphism. \texttt{src\_alts} is the list of picked coangles, and
\texttt{tgt\_alts} is the list of picked angles.

\hypertarget{add-a-terminal-morphism}{%
\subsection{Add a Terminal Morphism}\label{add-a-terminal-morphism}}

Above we discuss how to add a non-terminal morphism \emph{for two
reachable objects}. It is not the complete inverse of ``remove a
morphism'' because there are two special cases: removing an initial or
terminal morphism will result in the object being removed, which also
drop a lot of information, so that cannot then be directly added back.
The reverse processes of these two cases cannot be done by this method.

In categorical perspective, to add a terminal morphism
\(\phi : F \to \mathbb{U}\), one should also add an object \(F\) and its
incoming morphisms. An incoming morphism, say \(\psi : S \to F\), and
the added terminal morphism \(\phi\) should be composable, and the
result, denoted as \(\phi \circ' \psi\), should be assigned to an unique
terminal morphism \(\zeta : S \to \mathbb{U}\). The compositions between
added incoming morphisms \(\psi : S \to F\) and existing morphisms, say
\(\xi : V \to S\), also need to be given, the result, denoted as
\(\psi \circ'' \xi\), should be assigned to another incoming morphism
\(\psi' : V \to F\). Also, the new composition rules \(\circ''\) and
\(\circ'\) should respect to original ones: for two compositions
\(\psi' = \psi \circ'' \xi\) and \(\zeta = \phi \circ' \psi\), there is
\(\zeta' = \phi \circ' \psi'\) for \(\zeta' = \zeta \circ \xi\).

Consider an added incoming morphism \(\psi : S \to F\), for any two
parallel incoming morphisms \(\xi, \xi' : V \to S\), one should decide
\(\psi \circ'' \xi\) and \(\psi \circ'' \xi'\) should be equivalent or
not. As a reverse process of removing terminal morphism, such
information has been dropped by removing terminal morphism. To minimize
input information, we choose the most trivial rule: let
\(\psi \circ'' \xi\) and \(\psi \circ'' \xi'\) be assigned to different
morphisms if \(\xi \neq \xi'\). Where we further assume that the source
objects of all added incoming morphisms have a unique maximum \(S\), so
that all other incoming morphisms can be unqiuely derived by the
incoming morphism \(\psi : S \to F\). In this setup, the new composition
rules already respect to original ones. This operation can also be seen
as inserting an object in the middle of a terminal morphism, such that
it can be decomposed into two added morphisms.

Utilizing this simplified operation, the reverse process of removing a
terminal morphism now can be built. First, insert an object in the
middle of a terminal morphism, which will not introduce additional
structure. Second, repeat the first step, and merge all added objects by
merging their initial morphisms (see
\protect\hyperlink{merge-non-terminal-morphisms}{Merge Non-Terminal
Morphisms}), so that all incoming morphisms of all added objects are
disjointly unioned together. And third, merge some incoming morphisms of
added objects (see
\protect\hyperlink{merge-non-terminal-morphisms}{Merge Non-Terminal
Morphisms}), so that the additional structure can be encoded.

In geometrical perspective, it corresponds to ``add a trivial facet'',
which is a reverse process of ``removing a trivial facet''. A facet is
said to be trivial if it has no superfacet and has only one direct
subfacet, and the incidence relation is directly inherited from this
subfacet. This is like extruding a facet, which make a superfacet on top
of it. No subfacet will be glued after extruding (no new incidence
relation will be created).

\begin{figure}
\hypertarget{fig:add-leafnode}{%
\centering
\includegraphics[width=1\textwidth,height=\textheight]{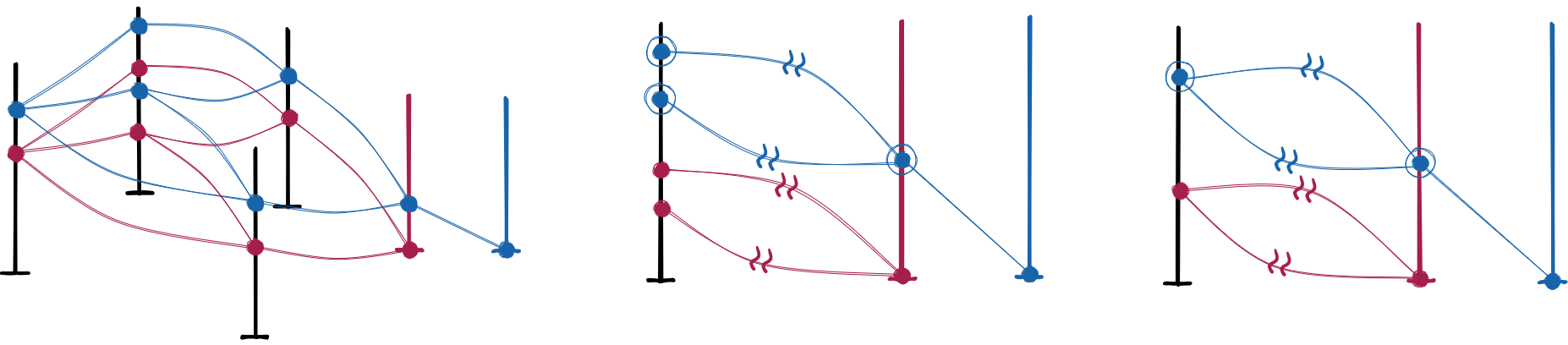}
\caption{Add a leaf node (blue pole) to a given node (red pole). The
left diagram shows that all added wires (blue lines) should have the
same shape to the supporting base wires (red lines). The two diagrams on
the right shows the rules: if two base wires (red lines) start at the
different/same points, the added wires (blue lines) should also start at
the different/same points.}\label{fig:add-leafnode}
}
\end{figure}

In program, it is a process to add a leaf node to a node. A symbol is
also added to the node, which is connected to the added leaf node. The
wires to the leaf node should also be added for every ancestor nodes,
which should have the same shape to the base wires of the node (see
Figure \ref{fig:add-leafnode}). It is implemented as a function
\texttt{addLeafNode} with parameters
\VERB|\NormalTok{src,}\OtherTok{ sym ::} \DataTypeTok{Symbol}| and
\VERB|\OtherTok{inserter ::}\NormalTok{ (}\DataTypeTok{Symbol}\NormalTok{, }\DataTypeTok{Symbol}\NormalTok{) }\OtherTok{{-}\textgreater{}} \DataTypeTok{Symbol}|.
\texttt{src} indicates an object whose terminal morphism will be
interpolated, and \texttt{(src,\ sym)} will indicate the only morphism
from such object to the inserted object. For all incoming morphisms of
the object \texttt{src}, say \texttt{(s1,\ s2)}, the pair of symbol
\texttt{(s1,\ inserter\ (s1,\ s2))} will indicate the incoming morphism
of the inserted object with the same source object.

\hypertarget{add-an-initial-morphism}{%
\subsection{Add an Initial Morphism}\label{add-an-initial-morphism}}

The opposite of above operation is to add an initial morphism. To add an
initial morphism \(\phi : \varnothing \to F\), one should also add an
object \(F\) and its outgoing morphisms. An outgoing morphism, say
\(\psi : F \to S\), and the added initial morphism \(\phi\) should be
composable, and the result, denoted as \(\psi \circ' \phi\), should be
assigned to a unique initial morphism \(\zeta : \varnothing \to S\). The
compositions between added outgoing morphisms \(\phi : F \to S\) and
existing morphisms, say \(\zeta : S \to V\), also need to be given, the
result, denoted as \(\xi \circ'' \psi\), should be assigned to another
outgoing morphism \(\psi' : F \to V\). Also, the new composition rules
\(\circ''\) and \(\circ'\) should respect to original ones: for two
compositions \(\psi' = \xi \circ'' \psi\) and
\(\zeta = \psi \circ' \phi\), there is \(\zeta' = \psi' \circ' \phi\)
for \(\zeta' = \xi \circ \zeta\).

Similarly, we only consider the most trivial case. There is a minimal
object \(S\) such that the added outgoing morphism \(\psi : F \to S\)
uniquely determines the other outgoing morphisms by composition:
\(\xi \circ'' \psi\) and \(\xi' \circ'' \psi\) are different outgoing
morphisms iff \(\xi \neq \xi'\). This operation can also be seen as
inserting an object in the middle of an initial morphism, such that it
can be decomposed into two added morphisms.

The reverse process of removing an initial morphism now can be built.
First, insert an object in the middle of an initial morphism. Second,
repeat the first step, and merge all added objects into one (see
\protect\hyperlink{merge-terminal-morphisms}{Merge Terminal Morphisms}),
so that all outgoing morphisms of all added objects are disjointly
unioned together. And third, merge some outgoing morphisms of added
objects (see \protect\hyperlink{merge-non-terminal-morphisms}{Merge
Non-Terminal Morphisms}), so that the additional structure can be
encoded.

\begin{figure}
\hypertarget{fig:insert-node-on-root}{%
\centering
\includegraphics[width=0.2\textwidth,height=\textheight]{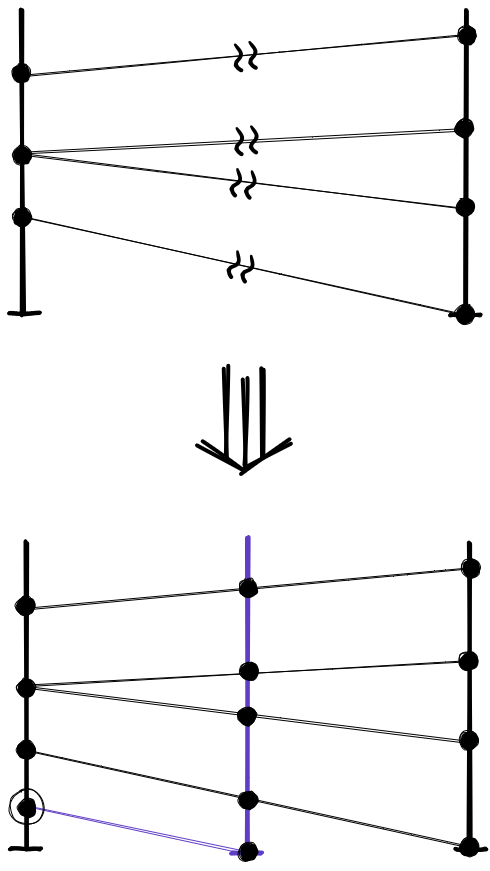}
\caption{Insert a node (purple pole) in the middle of an arrow. The
outgoing edge of the added node has one-to-one dictionary, and the
incoming edge has the same shape as the original arrow except the new
base wire (purple line).}\label{fig:insert-node-on-root}
}
\end{figure}

In geometrical perspective, it creates an boundaryless subfacet on a
facet, for example, creates a point on a plane, or creates a circle in a
box. The created subfacet is placed at the inner of the facet, so that
the incidence relations to its superfacets are directly inherited from
it.

In program, it is a process that add a node in the middle of an arrow
started at the root (see Figure \ref{fig:insert-node-on-root}). The
inserted node has only one edge, and its dictionary is one-to-one, so
there is no additional structure. A corresponding symbol should be added
to the root node, which connects to the base point of the added node. It
is implemented as a function \texttt{addParentNodeOnRoot} with
parameters
\VERB|\NormalTok{tgt,}\OtherTok{ sym ::} \DataTypeTok{Symbol}| and
\VERB|\OtherTok{mapping ::} \DataTypeTok{Dict}|, where \texttt{tgt}
indicates the initial morphism to be interpolated, and \texttt{sym} will
indicate the incoming morphism of the added object, and \texttt{mapping}
is the dictionary of the outgoing morphism of the added object.

\hypertarget{split-a-non-terminal-morphism}{%
\subsection{Split a Non-Terminal
Morphism}\label{split-a-non-terminal-morphism}}

\textbf{``Disconnect''} is used to separate disconnected components of a
disconnected facet, such ill facet may be produced after claiming
incidence relation. For example, after claim that a circle is covered by
a sphere, the sphere becomes disconnected: the sphere has been cut into
a cap and a cup. Local connectedness can also be manipulated by this
way.

The connectedness of a facet is manifested by identity of morphisms, so
in categorical perspective, it corresponds to splitting a morphism into
multiple parts, and each part represents each connected component. After
splitting the target morphism, one should also figure out how to modify
the multiplication table. The composition of the target morphism and
other morphisms will be duplicated for each splitted one. For the
composition that result in the target morphism, things are complicated.
We need to find all morphism chains of the target morphism, and
determine which chain will compose which splitted morphism. For example,
a sphere with a circle on it can be splitted into a cap and a cup; one
covers this circle in positively-oriented way, another in
negatively-oriented way. Splitting the maximum chains means breaking
equivalence relations, which is not always possible; one should find a
way to consistently break equivalence relations. All equivalence
relations to be broken should be a minimal equivalence relation: for
equivalence relation
\(\phi_1 \circ \dots \circ \phi_n = \phi_1' \circ \dots \circ \phi_t'\)
there is no non-trivial subpath decomposition
\((\phi_1 \circ \dots \circ \phi_m) \circ (\phi_{m+1} \circ \dots \circ \phi_n) = (\phi_1' \circ \dots \circ \phi_s') \circ (\phi_{s+1}' \circ \dots \circ \phi_t')\)
such that
\(\phi_1 \circ \dots \circ \phi_m = \phi_1' \circ \dots \circ \phi_s'\)
and
\(\phi_{m+1} \circ \dots \circ \phi_n = \phi_{s+1}' \circ \dots \circ \phi_t'\).
If one try to break some relations aren't minimal equivalence relations,
one of sub-relations should also be broken. In this case, one should
break this sub-relation first in the same way. So it is limited that
broken equivalence relations should be minimal. The constraint of
minimal equivalence relations just corresponds to the condition of
linked objects of the section category of the target morphism, in this
point of view, this process just separates out splittable parts in a
category.

\begin{figure}
\hypertarget{fig:split-wires}{%
\centering
\includegraphics[width=0.7\textwidth,height=\textheight]{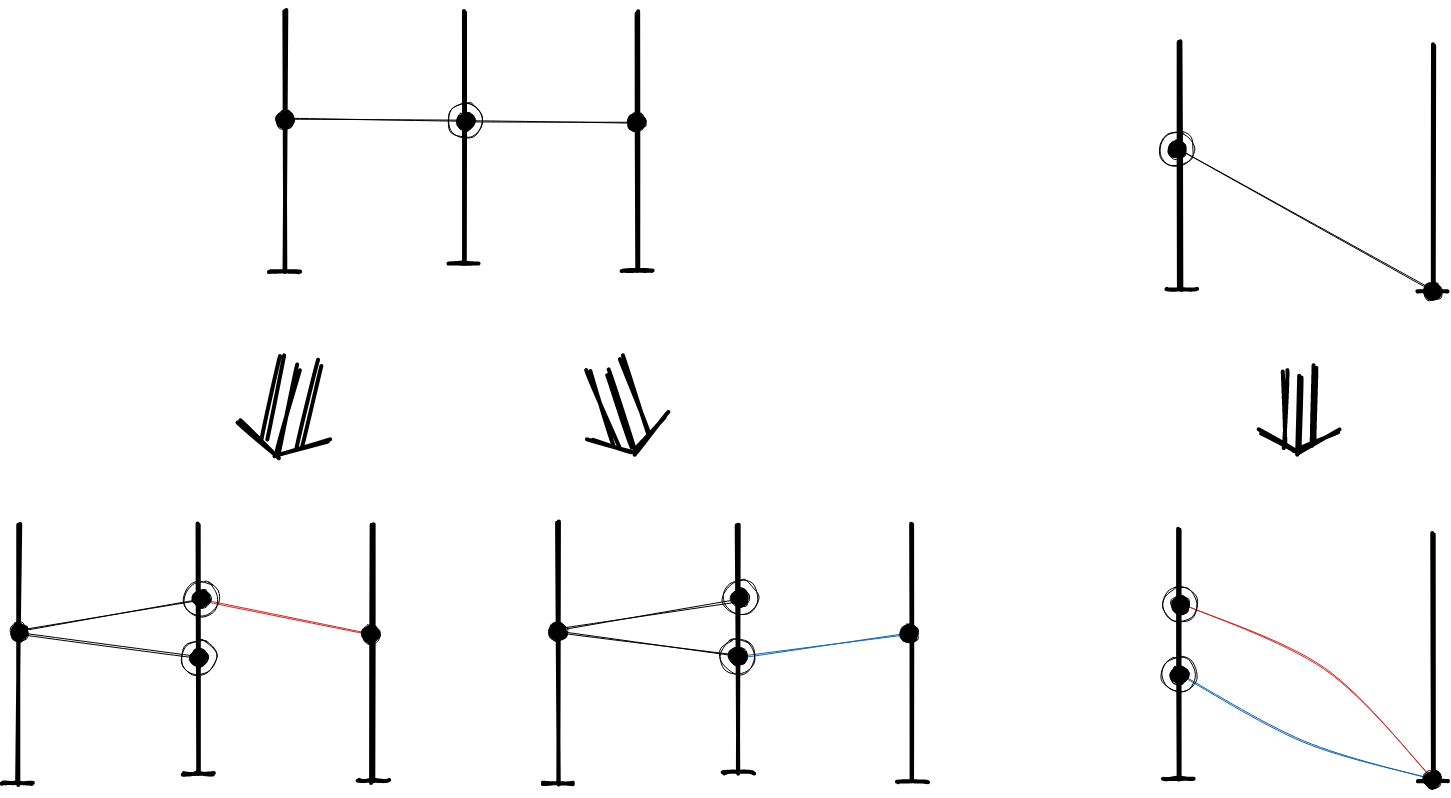}
\caption{Split a symbol (circled point) on a node. Left one shows that
splitting symbols results in splitting of incoming wires and choosing a
side for outgoing wires (red line or blue line). Right one shows that
splitting a symbol with a direct edge to the referenced node results in
duplicating this edge (red line and blue line).}\label{fig:split-wires}
}
\end{figure}

In program, this process splits a symbol on a node into two, and keep
all other unrelated wires unchanged. Focus on the wires passing through
the point to be splitted, the incoming wire should be splitted in two,
and the outgoing wire should choose to connect to one of the splitted
points (see left one of Figure \ref{fig:split-wires}). There is a
special case: if outgoing wire directly connects to a base point, this
edge will be duplicated so that there are two base wires that connect to
two splitted points respectively (see right one of Figure
\ref{fig:split-wires}).

\begin{figure}
\hypertarget{fig:split-constraints}{%
\centering
\includegraphics[width=0.7\textwidth,height=\textheight]{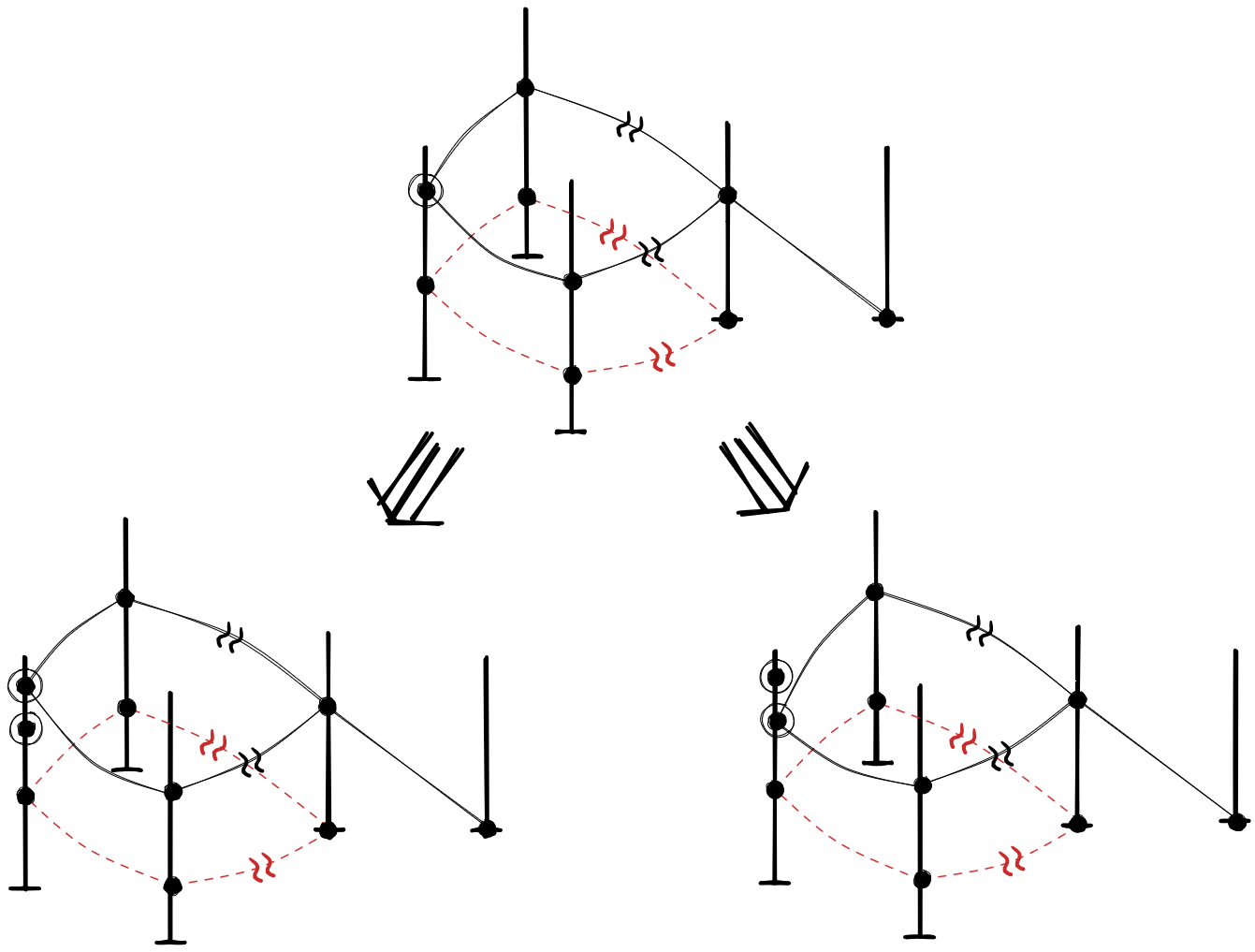}
\caption{Constraints for splitting symbol. Two outgoing wires which are
supported by a circle of base wires (red dashed lines) should connect to
the same splitted point (circled point).}\label{fig:split-constraints}
}
\end{figure}

The new dictionaries of incoming edges should map two splitted symbols
to the same symbol as before, and the new dictionaries of outgoing edges
should maps to one of the splitted symbols: it should be decided where
the mapping to the old symbol should now maps to, and such choices
aren't arbitrary. Considering two outgoing wires starting at the point
to be splitted, they should be modified to start at the same splitted
symbol if they are supported by a circle of two base wires (see Figure
\ref{fig:split-constraints}). Since such base wires are not modified,
the supportivity ensures that they should start at the same point. This
constraint is the same as the condition of linked objects.

A function \texttt{partitionPrefix} is defined to determine splittable
parts, which has a parameter
\VERB|\OtherTok{tgt ::} \DataTypeTok{Symbol}|, indicating the morphism
to partition. This function searches for all \(3\)-chains of given
morphism, in which the first and the third arrows are edges. For each
\(3\)-chain, its two direct subchains indicate two linked minimal and
maximal objects. The return value is a list of groups of tuples of
symbols, each group contains all symbols on child nodes that should be
mapped to the same splitted symbol. The main process is implemented as a
function \texttt{splitSymbol}, which has two parameters
\VERB|\NormalTok{(src, tgt)}\OtherTok{ ::}\NormalTok{ (}\DataTypeTok{Symbol}\NormalTok{, }\DataTypeTok{Symbol}\NormalTok{)}|
and
\VERB|\OtherTok{partition ::}\NormalTok{ [(}\DataTypeTok{Symbol}\NormalTok{, [(}\DataTypeTok{Symbol}\NormalTok{, }\DataTypeTok{Symbol}\NormalTok{)])]}|.
\texttt{(src,\ sym)} for all
\texttt{(sym,\ group)\ \textless{}-\ partition} indicates a splitted
morphism, and the morphism chains in the group \texttt{group} will
compose to this morphism. An splitted morphism which no morphism chains
compose to is nondecomposable.

\hypertarget{split-a-terminal-morphismcategory}{%
\subsection{Split a Terminal
Morphism/Category}\label{split-a-terminal-morphismcategory}}

Splitting a non-terminal morphism corresponds to disconnect a facet or a
vertex figure, while splitting a terminal morphism, which leads to the
source object being splitted, corresponds to split a facet from two
sides, so we call this process \textbf{``split''}. For example, a point
covered by two segments in two directions can be splitted into two
points, so that they are covered by two segments respectively. It is
useful when one needs to separate two disconnected facets with shared
boundaries. Complex shared boundaries can be splitted from up to down
using ``split''. For example, to split two sticked cubes, one needs to
split the shared facet first, then split shared edges and shared
vertices. After that, one can split this category into two, and rotating
or moving them respectively now works. Splitting the terminal morphism
of the initial object is a special case, which is just splitting a
category into multiple categories. this is the reverse process of
merging categories.

In categorical perspective, splitting a terminal morphism
\(\phi : F \to \mathbb{U}\) into \(N\) parts is special since the object
\(F\) is also splitted. Such process can be further decomposed into:
duplicate all incoming morphisms of object \(F\), then split initial
morphism \(\phi' : \varnothing \to F\), terminal morphism
\(\phi : F \to \mathbb{U}\) and object \(F\) simultaneously. To
duplicate all incoming morphisms, one should start with minimal ones, so
that longer ones can be duplicated coherently. Denote the duplications
of an incoming morphism \(\psi' : P \to F\) as
\((\psi_n' : P \to F)_{n = 1 \sim N}\). ``Duplicate'' means they behave
the same up to indices, that is,
\(\xi \circ \psi_n' = \xi \circ \psi_m'\) for all \(\xi : F \to G\) and
\(\psi_n' \circ \zeta = \psi_n'' \iff \psi_m' \circ \zeta = \psi_m''\)
for all \(\zeta : Q \to P\). To split initial and terminal morphisms and
the given object simultaneously, the target object of incoming morphisms
and the source object of outgoing morphisms will change. Let's say the
outgoing morphism \(\xi_m : F \to G\), which is assigned to be splitted
into \(m\)-th group, now becomes \(\tilde{\xi}_m : F_m \to G\), and
incoming morphisms \((\psi_n' : P \to F)_{n = 1 \sim N}\) now become
\((\tilde{\psi}_n' : P \to F_n)_{n = 1 \sim N}\). For any pair of an
incoming morphism \(\tilde{\psi}_n'\) and an outgoing morphism
\(\tilde{\xi}_m\) with \(n \neq m\), the composition between them is no
longer possible, so such case will be removed from the multiplication
table. Before this step, there is a morphism
\(\eta = \xi_m \circ \psi_n'\), which will not be removed because for
any incoming morphism \(\tilde{\psi}_n' : P \to F_n\), there is
\(\tilde{\psi}_m' : P \to F_m\) such that
\(\tilde{\xi}_m \circ \tilde{\psi}_m' = \eta\). Splitting an initial
morphism also satisfies the same statement in an opposite way.

\begin{figure}
\hypertarget{fig:split-node}{%
\centering
\includegraphics[width=0.7\textwidth,height=\textheight]{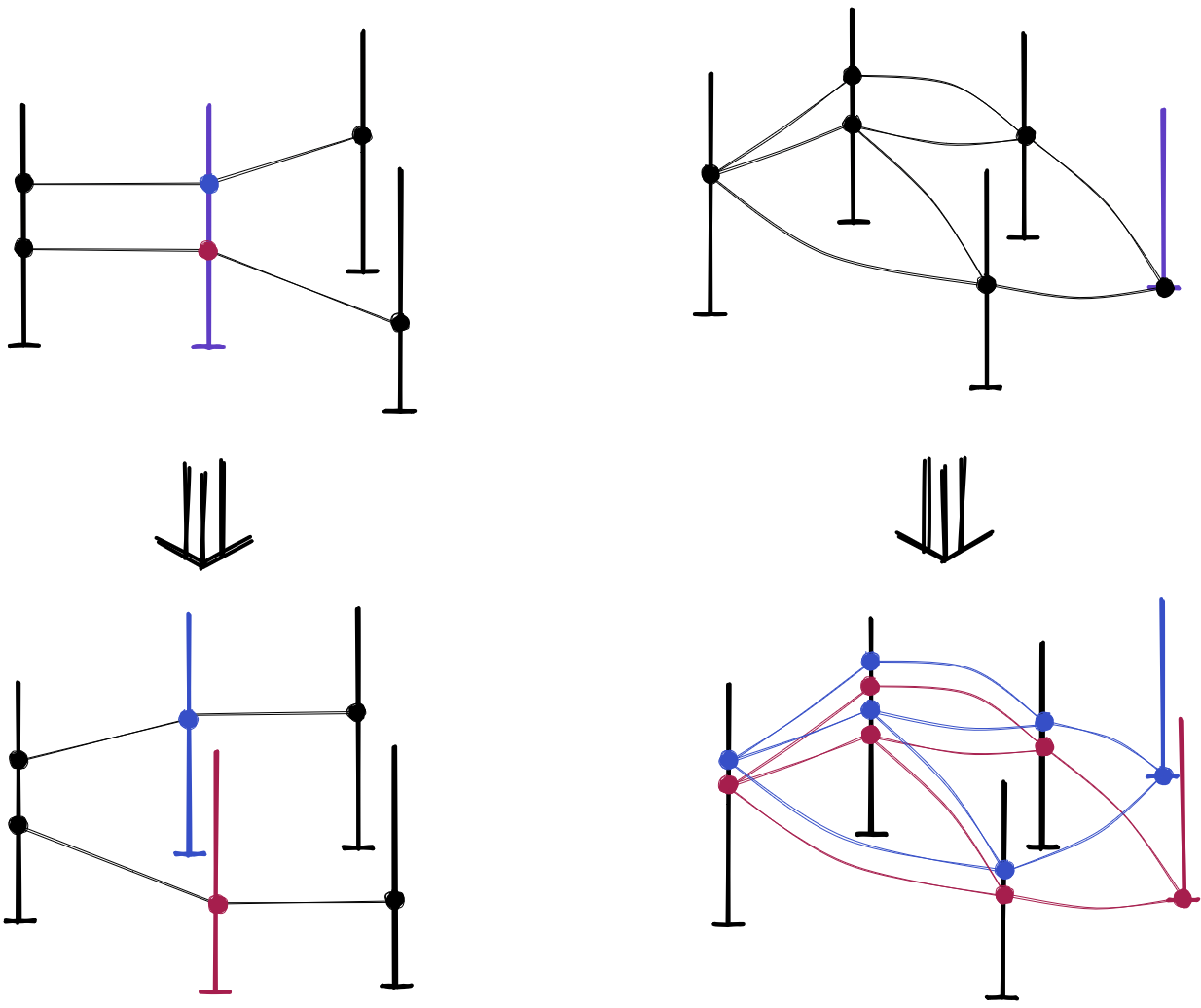}
\caption{Split a node (purple pole). Left one shows symbols on the node
(red point and blue point) will be distributed to each part (red pole
and blue pole), as will the connected wires. Right one shows the points
connecting to the base point of the node will be duplicated (red points
and blue points).}\label{fig:split-node}
}
\end{figure}

\begin{figure}
\hypertarget{fig:split-node-constraint}{%
\centering
\includegraphics[width=0.4\textwidth,height=\textheight]{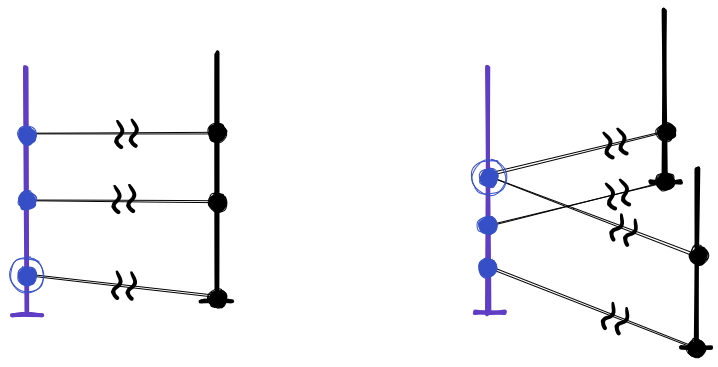}
\caption{Constraints for splitting a node. It shows that the symbols
(normal points) support (right one) or are supported by (left one) the
same symbol (circled points) should be distributed to the same
part.}\label{fig:split-node-constraint}
}
\end{figure}

In program, this process splits a node into multiple parts. The symbols
on this node (except the base symbol) should be distributed to each
part, as should the wires connected to them (see left one of Figure
\ref{fig:split-node}). The symbols support or are supported by the same
symbol should be distributed to the same part (see Figure
\ref{fig:split-node-constraint}). All wires that connect to the base
symbol of this node should be duplicated, so all passing points will
also be duplicated (see right one of Figure \ref{fig:split-node}). For
the special case of splitting a category, it should be defined in
another function since it has different return type. Because it is
equivalent in the categorical perspective, the criteria is the same.

To determine splittable parts, an utility function
\texttt{partitionSymbols} is also defined, which is similar to
\texttt{partitionPrefix}. The function \texttt{partitionSymbols} finds
splittable groups of symbols by applying union find algorithm to all
values of dictionaries of edges. The process of splitting a category is
defined as a function \texttt{splitRootNode} with a parameter
\VERB|\OtherTok{partition ::}\NormalTok{ [[}\DataTypeTok{Symbol}\NormalTok{]]}|,
which are groups of symbols representing each splitted category.
Splitted categories are built just by separating the edges of the root
via \texttt{partition}. The process of splitting an object is defined as
a function \texttt{splitNode} with parameters
\VERB|\OtherTok{tgt ::} \DataTypeTok{Symbol}|, the symbol of the object
to split, and
\VERB|\OtherTok{partition ::}\NormalTok{ [((}\DataTypeTok{Symbol}\NormalTok{, }\DataTypeTok{Symbol}\NormalTok{) }\OtherTok{{-}\textgreater{}} \DataTypeTok{Symbol}\NormalTok{, [}\DataTypeTok{Symbol}\NormalTok{])]}|.
For all incoming morphisms of the object to split, say
\texttt{(s1,\ s2)}, the pair of symbol
\VERB|\NormalTok{(s1, splitter (s1, s2))}| for
\texttt{(splitter,\ \_)\ \textless{}-\ partition} will indicate the
incoming morphism of splitted object with the same source object.

\hypertarget{merge-non-terminal-morphisms}{%
\subsection{Merge Non-Terminal
Morphisms}\label{merge-non-terminal-morphisms}}

The reversed process of ``disconnect'' is \textbf{``connect''}. It is
usually used to union two disjoint facets. This is needed when, for
example, merging two sticked squares into one rectangle: before remove
the boundary between them, one should union the area of two squares. In
that moment, the merged geometric object becomes disconnected. The
facets to be merged should be covered by the same facets. For example,
you cannot merge two sticked squares which are faces of two different
cubes; you should first merge the two cubes.

In categorical perspective, this process merges multiple non-terminal
morphisms into one. Merging two morphisms \(\phi_{n0}\) and
\(\phi_{m0}'\) means letting \(\phi_{n0} = \phi_{m0}'\), which implies
equivalence relations
\(\phi_n \circ \dots \circ \phi_1 \circ \phi_0 = \phi_m' \circ \dots \circ \phi_1' \circ \phi_0'\)
between their morphism chains
\(\langle \phi_n, \dots, \phi_2, \phi_1 \rangle\) and
\(\langle \phi_m', \dots, \phi_2', \phi_1' \rangle\). To merge two
morphisms \(\phi : S \to F\) and \(\phi' : S \to F\), if there is a
morphism \(\psi : P \to S\), the equivalence relation
\(\phi \circ \psi = \phi' \circ \psi\) should already exist. Similarly,
if there is a morphism \(\psi : F \to V\), the equivalence relation
\(\psi \circ \phi = \psi \circ \phi'\) should already exist. When
merging two initial morphisms \(\phi : \varnothing \to F\) and
\(\phi' : \varnothing \to F'\), the target objects of morphisms are not
the same. In this case, one should find an \textbf{upper isomorphism} of
their upper categories, which is an isomorphism
\(\mu : \mathcal{F} \operatorname{\uparrow} F' \to \mathcal{F} \operatorname{\uparrow} F\)
such that their downward functors are equivalent under it:
\(F^\downarrow \circ \mu = F'^\downarrow\).

\begin{figure}
\hypertarget{fig:merge-symbols-constraint}{%
\centering
\includegraphics[width=0.3\textwidth,height=\textheight]{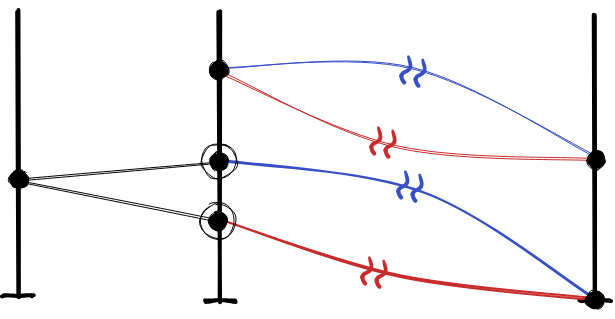}
\caption{Constraint for merging two symbols (circled points). The
incoming wires (black lines) should connect to the same point. The wires
(upper blue line and upper red line) supported by the base wires of this
two symbols (lower blue line and lower red line) should have the same
startpoint and endpoint.}\label{fig:merge-symbols-constraint}
}
\end{figure}

In program, this process merges multiple symbols on a node, and keep all
other unrelated wires unchanged. The incoming edge of this node should
maps the symbols to be merged to the same point. The nodes referenced by
these symbols should be the same, and their dictionaries should be the
same except the base one, otherwise supportivity will be violated (see
Figure \ref{fig:merge-symbols-constraint}). If their dictionaries map
one symbol to two different symbols, user should first merge those
symbols. When merging non-initial morphisms, there is no need to check
whether they have the same target node, since it is already implied by
another requirement. But for merging initial morphisms, this is
important. To merge initial morphisms, the target nodes should be
equivalent in data structure. If they are equivalent in category but not
in data structure, they should be unified via \texttt{relabel} and
\texttt{rewire}. Only target nodes are needed to be relabeled and
rewired, since their descendants are already the same, as implied by
another requirement.

The process of merging morphisms is implemented as a function
\texttt{mergeSymbols} with parameters
\VERB|\NormalTok{(src, tgts)}\OtherTok{ ::}\NormalTok{ (}\DataTypeTok{Symbol}\NormalTok{, [}\DataTypeTok{Symbol}\NormalTok{])}|,
indicating morphisms to merge, and merged symbol
\texttt{sym\ ::\ Symbol}, so \texttt{(src,\ sym)} will be indicate the
merged morphism. When merging initial morphisms, it will checks if the
structures of the target nodes are the same. Users should unify target
nodes before merging.

\hypertarget{merge-terminal-morphisms}{%
\subsection{Merge Terminal Morphisms}\label{merge-terminal-morphisms}}

In the previous subsection, we discuss the process of merging
non-terminal morphisms called ``connect'', which is the reverse process
of ``disconnect'', while merging terminal morphisms is called
\textbf{``merge''}, which is the reverse process of ``split''. It can be
used to merge facets. For example, stick two squares by merging their
edges, so that this edge now is covered by the two squares. The facets
to be merged should cover the same facets. In the above case, before
merging edges, their vertices should be merged first.

A categorical perspective can be obtained by reversing the discussion of
\protect\hyperlink{split-a-terminal-morphismcategory}{splitting a
terminal morphism}. Merging \(N\) terminal morphisms
\((\phi_n : F_n \to \mathbb{U})_{n = 1 \sim N}\) is special since their
source objects are different. Such process can be further decomposed
into: merge initial morphisms
\((\phi_n' : \varnothing \to F_n)_{n = 1 \sim N}\), terminal morphisms
\((\phi_n : F_n \to \mathbb{U})_{n = 1 \sim N}\) and objects
\((F_n)_{n = 1 \sim N}\) simultaneously, then un-duplicate all incoming
morphisms of the merged object. To merge initial and terminal morphisms
and objects simultaneously, the target object of incoming morphisms and
the source object of outgoing morphisms will change. Let's say the
outgoing morphism \(\xi_m : F_m \to G\) now becomes
\(\tilde{\xi}_m : F \to G\), and the incoming morphism
\(\psi_n' : P \to F_n\) now becomes \(\tilde{\psi}_n' : P \to F\).
Incoming morphisms have been grouped into
\((\tilde{\psi}_n' : P \to F)_{n = 1 \sim N}\) according to certain
properties, such that the second step can be done. For any pair of an
incoming morphism \(\tilde{\psi}_n'\) and an outgoing morphism
\(\tilde{\xi}_m\) with \(n \neq m\), the composition between them is now
possible, so such case should be added to the multiplication table. The
result of \(\tilde{\xi}_m \circ \tilde{\psi}_n'\) should be set as
\(\tilde{\xi}_m \circ \tilde{\psi}_m'\), where \(\tilde{\psi}_n'\) and
\(\tilde{\psi}_m'\) are in the same group. To un-duplicate all incoming
morphisms, one should start with maximal ones, so that the shorter ones
can be un-duplicated coherently. Each group of incoming morphisms
\((\tilde{\psi}_n' : P \to F)_{n = 1 \sim N}\) will be identified as the
same morphism, then be denoted as \(\tilde{\psi}' : P \to F\), which is
called an un-duplication. They can be ``un-duplicated'' only if they
behave the same up to indices, that is,
\(\xi \circ \tilde{\psi}_n' = \xi \circ \tilde{\psi}_m'\) for all
\(\xi : F \to G\) and
\(\tilde{\psi}_n' \circ \zeta = \tilde{\psi}_n'' \iff \tilde{\psi}_m' \circ \zeta = \tilde{\psi}_m''\)
for all \(\zeta : Q \to P\), where \(\tilde{\psi}_n' : P \to F\) and
\(\tilde{\psi}_m' : P \to F\) are in the same group,
\(\tilde{\psi}_n'' : Q \to F\) and \(\tilde{\psi}_m'' : Q \to F\) are in
the same group.

Conclude above discussion, to merge terminal morphisms
\((\phi_n : F_n \to \mathbb{U})_{n = 1 \sim N}\), objects
\((F_n)_{n = 1 \sim N}\) should have the same lower closure on the
induced poset except upper bounds. Furthermore, their incoming morphisms
should be grouped into \((\psi_n' : P \to F_n)_{n = 1 \sim N}\) such
that they behave the same up to indices:
\(\phi_n \circ \psi_n' = \phi_m \circ \psi_m'\) and
\(\psi_n' \circ \zeta = \psi_n'' \iff \psi_m' \circ \zeta = \psi_m''\)
for all \(\zeta : Q \to P\) and \((\psi_n' : P \to F_n)_{n = 1 \sim N}\)
and \((\psi_n'' : Q \to F_n)_{n = 1 \sim N}\). This requirement is
equivalent to find \textbf{lower isomorphisms} between them, which are
isomorphisms between their lower categories
\(\mu_{nm} : \mathcal{F} \operatorname{\downarrow} F_m \to \mathcal{F} \operatorname{\downarrow} F_n\)
such that their upward functors are equivalent under it:
\(F_n^\uparrow \circ \mu_{nm} = F_m^\uparrow\).

\begin{figure}
\hypertarget{fig:merge-nodes}{%
\centering
\includegraphics[width=0.5\textwidth,height=\textheight]{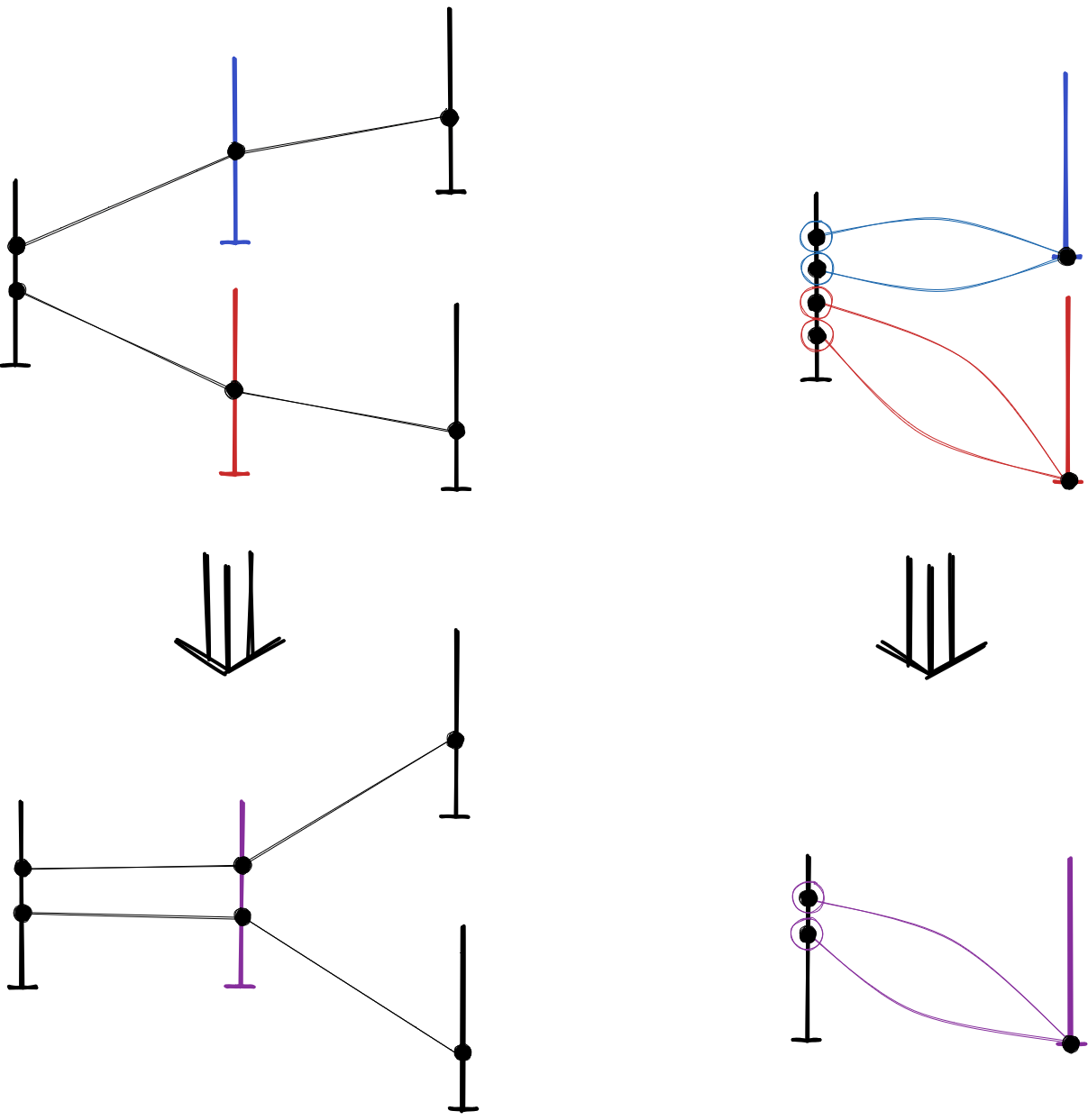}
\caption{Merge two nodes (blue pole and red pole). Left one shows wires
passing through the nodes being merged (blue pole and red pole) are now
passing through the merged node (purple pole) without crossing. Right
one shows the base points of the nodes (normal points) should be merged
into one point, so the base wires connected to these points (blue lines
and red lines) should be merged coherently, which leads to some points
(blue circled points and red circled points) to be
merged.}\label{fig:merge-nodes}
}
\end{figure}

\begin{figure}
\hypertarget{fig:merge-invalid}{%
\centering
\includegraphics[width=0.2\textwidth,height=\textheight]{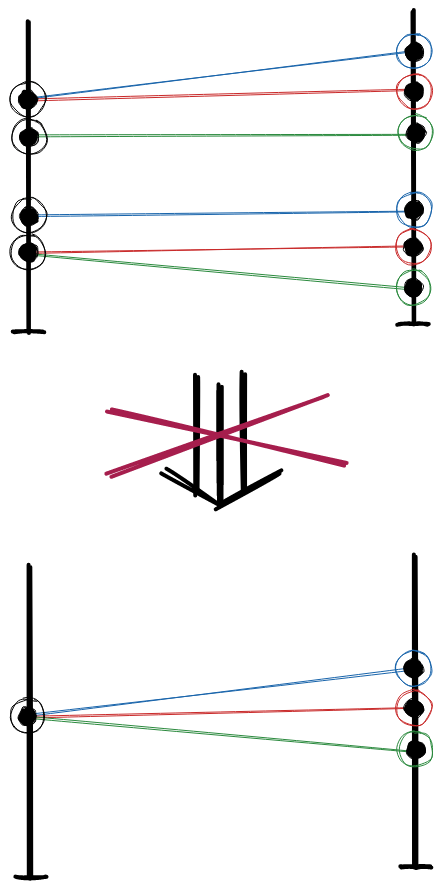}
\caption{An invalid case happens when merging symbols referencing the
nodes being merged. To merge two triples of symbols (blue, red and green
circled points), the shape of incoming wires should be the same (blue,
red and green lines), otherwise the result will have different
shape.}\label{fig:merge-invalid}
}
\end{figure}

In program, this process merges multiple nodes into one node. All
symbols (except base symbols) on the nodes being merged are disjointly
unioned. Wires connected to these symbols will remain the same. But the
wires connected to base symbols should be modified since they are now
merged into one base symbol (see left one of Figure
\ref{fig:merge-nodes}). Consider the common parent of the nodes to be
merged, which may have multiple edges between them (see right one of
Figure \ref{fig:merge-nodes}). To merge nodes, these edges should also
be merged. It should be decided which base wire should be merged into
which. It causes some symbols on the parent node to be merged. All
symbols referencing the nodes being merged also need to be merged
coherently. ``Coherent'' means the shape of base wires should remain the
same after merging. Note that incorrectly merging incoming edges will
result in inconsistent situations. For example, consider two triples of
symbols on a node, say \texttt{(1,\ 2,\ 3)} and
\texttt{(1\textquotesingle{},\ 2\textquotesingle{},\ 3\textquotesingle{})},
and assume there are some incoming wires:
\texttt{\{1→a;\ 2→a;\ 3→b;\ 1\textquotesingle{}→a\textquotesingle{};\ 2\textquotesingle{}→b\textquotesingle{};\ 3\textquotesingle{}→b\textquotesingle{}\}}.
In this case, merging \texttt{(1,\ 2,\ 3)} and
\texttt{(1\textquotesingle{},\ 2\textquotesingle{},\ 3\textquotesingle{})}
will lead to merging all four symbols \texttt{a}, \texttt{b},
\texttt{a\textquotesingle{}}, \texttt{b\textquotesingle{}}, which
changes the configuration of the wires (see Figure
\ref{fig:merge-invalid}).

It is implemented as a function \texttt{mergeNodes} with parameters
\VERB|\OtherTok{tgts\_suffix ::}\NormalTok{ [(}\DataTypeTok{Symbol}\NormalTok{, [(}\DataTypeTok{Symbol}\NormalTok{, }\DataTypeTok{Symbol}\NormalTok{)])]}|
and
\VERB|\OtherTok{merger ::}\NormalTok{ (}\DataTypeTok{Symbol}\NormalTok{, [}\DataTypeTok{Symbol}\NormalTok{]) }\OtherTok{{-}\textgreater{}} \DataTypeTok{Symbol}|,
where \texttt{tgts\_suffix} contains nodes to merge and the
corresponding nondecomposable incoming edges, and \texttt{merger} is the
function to merge symbols on all ancestor nodes. All incoming morphisms
of these objects, say \texttt{(s,\ {[}r1,\ r2,\ ...{]})}, will be merged
into the morphism indicated by pair of symbol
\texttt{(s,\ merger\ (s,\ {[}r1,\ r2,\ ...{]}))}. The nondecomposable
incoming edges of the nodes to merge will be paired up by function
\texttt{zipSuffix} according to the keys.

\hypertarget{summary}{%
\section{Summary}\label{summary}}

Above we discussed fundamental operations in categorical level, which
can be classified as:

\begin{enumerate}
\def\labelenumi{\arabic{enumi}.}
\tightlist
\item
  \emph{remove}:
  \protect\hyperlink{remove-a-non-terminal-morphism}{removing a
  non-terminal morphism},
  \protect\hyperlink{remove-a-terminal-morphism}{removing a terminal
  morphism}.
\item
  \emph{add}: \protect\hyperlink{add-a-non-terminal-morphism}{adding a
  non-terminal morphism},
  \protect\hyperlink{add-a-terminal-morphism}{adding a terminal
  morphism} and \protect\hyperlink{add-an-initial-morphism}{adding an
  initial morphism}.
\item
  \emph{split}:
  \protect\hyperlink{split-a-non-terminal-morphism}{splitting a
  non-terminal morphism} and
  \protect\hyperlink{split-a-terminal-morphismcategory}{splitting a
  terminal morphism/category}.
\item
  \emph{merge}: \protect\hyperlink{merge-non-terminal-morphisms}{merging
  non-terminal morphisms},
  \protect\hyperlink{merge-terminal-morphisms}{merging terminal
  morphisms} and \protect\hyperlink{merge-categories}{merging
  categories}.
\end{enumerate}

These operations are fundamental because they directly manipulate
morphisms, like we do with graphs. Fundamental operations are not
simple, as some operations require multiple steps to complete. For
example, to add a morphism, it is required to analyze the situation
first then select required options, which are not arbitrary and may
fail; user should understand the mechanism behind it. Since initial and
terminal morphisms are special, operations of them are more complicated
and can be decomposed into multiple operations. For example, removing an
initial or terminal morphism can be decomposed into removing all related
morphisms then splitting category, which make adding an initial or
terminal morphism very complicated as a reverse process. Splitting
initial or terminal morphisms also can be decomposed into splitting all
related morphisms then splitting on duplications. A similar
decomposition can be applied to merge initial or terminal morphisms,
while should give isomorphisms between the objects to be merged. Noting
that adding/merging initial or terminal morphisms is much more
complicated than removing/splitting initial or terminal morphisms, and
the complexity comes from the huge information changes during these
operations.

In different perspectives, the similarity between operations are
different. In categorical perspectives, operations on initial and
terminal morphisms are very different from operations on non-initial
non-terminal morphisms. In programming perspectives, however, dealing
with initial morphisms and non-terminal morphisms are almost the same,
but not the same as dealing with terminal morphisms. The difference lies
in the way the BAC stores the information. A node in BAC contains the
information of the upper closure of an object, which constitutes a upper
category. This directional structure makes it lose the duality of
categories. Instead, it is more suitable for describing incidence
structures, thus making the interpretation of geometric objects more
intuitive. These complementary strengths drive investigations into
fundamental operations, and also provide a more stereoscopic approach to
analyze BAC.

There are some functions in the library we didn't mention.
\texttt{removeNode} and \texttt{addParentNode} are generalizations of
\texttt{removeLeafNode} and \texttt{addParentNodeOnRoot}, and such
generalizations are natural in the programming perspectives.
\texttt{duplicateNDSymbol} and \texttt{duplicateNode} are similar to
\texttt{splitNDSymbol} and \texttt{splitNode} in program.
\texttt{zipArrows} and \texttt{zipSuffixes} provide a way to validate
upper and lower isomorphisms and traverse BACs in parallel.

There are some operations we didn't implement, such as isomorphism,
section category. Various operations worked on cell complexes or
simplicial complexes, such as Cartesian product, join and connected sum,
also can be generalized. There are not necessary for building incidence
structures, so will not be discussed in this series of articles. In the
next article, we will develop a computational geometry system to
describe geometric objects consisting of curved facets in any dimension.

\printbibliography

\end{document}